\documentstyle[epsfig,12pt]{article}
 
\newlength{\dinwidth} 
\newlength{\dinmargin} 
\def\lapproxeq{\lower .7ex\hbox{$\;\stackrel{\textstyle 
<}{\sim}\;$}}
\def\gapproxeq{\lower .7ex\hbox{$\;\stackrel{\textstyle
>}{\sim}\;$}}

\def\msb{\overline{\rm MS}}
\def\GeV{{\rm GeV}}
\def\TeV{{\rm TeV}} 
 
\def\du{{d/u\uparrow}} 
\def\qup{{q\uparrow}} 
\def\qdown{{q\downarrow}} 
\def\sup{{s\uparrow}} 
\def\sdown{{s\downarrow}} 
\def\cup{{c\uparrow}} 
\def\cdown{{c\downarrow}} 
\def\gup{{g\uparrow}} 
\def\gdown{{g\downarrow}} 
\def\asup{{\alpha_S\uparrow\uparrow}}                                                   
\def\asdown{{\alpha_S\downarrow\downarrow}}                                                   
\def\as{{\alpha_S}}                                                   
\def\beq{\begin{equation}}                                                   
\def\eeq{\end{equation}}                                                   
\def\be{\begin{equation}} 
\def\ee{\end{equation}} 
\def\bea{\begin{eqnarray}}
\def\eea{\end{eqnarray}}
\def\pp{p \bar p}
                                                                                      
\begin{document} 
\titlepage                                                                                      
\begin{flushright}                                                                                      
DTP/99/64  \\
OUTP/9931P \\
RAL-TR-1999-047 \\
 hep-ph/9907231\\
July 1999
\end{flushright}                                                                                      
                                                                                      
\begin{center}
\vspace*{2cm}                                                                                      
{\Large \bf Parton Distributions and the LHC: $W$ and $Z$ Production} \\
\vspace*{1cm}
A.\ D.\ Martin$^a$, R.\ G.\ Roberts$^b$, W.\ J.\ Stirling$^{a,c}$ and R.\ S.\ Thorne$^d$ \\                                                                                      
                                                                                      
\vspace*{0.5cm} 
$^a \; $ {\it Department of Physics, University of Durham, 
Durham, DH1 3LE }\\                                                                  
$^b \; $ {\it Rutherford Appleton Laboratory, Chilton, 
Didcot, Oxon, OX11 0QX}\\                  
$^c \; $ {\it Department of Mathematical Sciences, University of Durham, Durham, DH1 3LE} \\  
$^d \; $ {\it Jesus College, University of Oxford, Oxford, OX1 3DW}                                                                                      
\end{center}                                                                                      
                                                                                      
\vspace*{1.5cm}                                                                                      
\begin{abstract}
$W$ and $Z$ bosons will be produced copiously at the LHC proton-proton
collider. We study the parton distribution dependence of the total
production cross sections and rapidity distributions, paying particular
attention to the uncertainties  arising from uncertainties in
the parton distributions themselves. Variations in the gluon, the strong
coupling, the sea quarks and the overall normalisation are shown to lead
to small but non-negligible variations in the cross section predictions.
Ultimately, therefore, the measurement of these cross sections will
provide a powerful cross check on our knowledge of parton distributions
and their evolution.
\end{abstract}                                                                                      
                                                                                      
\newpage                                                                                      
                                                                                      
\noindent{\large \bf 1. Introduction}                                                                                      

A precise knowledge of parton distribution functions (pdf's)
 is absolutely vital
 for reliable predictions for signal and background cross sections                             
 at the LHC. Uncertainties can arise both from the starting
distributions, obtained from a global fit to deep inelastic scattering
and other data,
 and from DGLAP evolution to the higher $Q^2$ scales typical
of LHC hard scattering processes.

There are several reasons why it is very difficult to derive `one sigma'
 errors on parton distributions of the form $f_i \pm \delta f_i$.
In the global fit there are complicated correlations between
a particular pdf at different
$x$ values, and between the different pdf flavours. To give an example: the
charm distribution is correlated with the gluon distribution, the gluon
distribution at low $x$ is correlated with the gluon at high $x$ via the
momentum sum rule, etc. Secondly, many of the uncertainties in the input
data or fitting procedure are not `true' errors in the statistical sense.
To give an example: the uncertainty in the high--$x$ gluon in the MRST
fits \cite{MRST} derives from a subjective assessment of the impact of `intrinsic
$k_T$' on the prompt photon cross section.

However, faced with the difficulties in trying to formulate {\it global}
pdf errors, what one  can do is make a detailed assessment
of the pdf uncertainty for a {\it particular}  cross section of interest,
by determining which partons contribute and at what $x$ and $Q^2$ values,
and then systematically tracing back to the data sets
that constrained the distributions in the global fit. Individual pdf sets
can then be constructed to reflect the uncertainty in the particular partons
determined by a particular data set (see Section~2).
The purpose of this paper is to show how this can be done for the
particular (test) case of $W$ and $Z$ production at the LHC.
The advantage of choosing these cross sections is that the theoretical
technology for calculating them is very robust. The total cross sections are
known to NNLO in QCD perturbation theory
\cite{WILLY}, and the input electroweak parameters
($M_{W,Z}$, weak couplings, etc.) are known to high accuracy. The complete
${\cal O}(\alpha)$ electroweak
radiative corrections are known, see for example
Ref.~\cite{ULI}.\footnote{Since the electroweak radiative corrections
depend on the nature of, and experimental cuts on, the final state
$W$ and $Z$ decay products~\cite{ULI}, we do not include them in our study.}
Theoretically, therefore, the main uncertainty derives from the input
pdfs and, to a lesser extent, $\alpha_S$.\footnote{The two are of course
correlated, see for example \cite{MRST}.}

Knowledge of the uncertainty on the theoretical prediction allows one to
assess the impact of an eventual experimental measurement. For example,
with what precision would a measurement of, say, $\sigma_W$ have to be made to
provide an additional constraint on the pdfs? We will not address
any experimental issues here. Some studies of the likely experimental
precision have already been  performed, see for example
Ref.~\cite{DITTMAR}, with encouraging results.

\begin{figure}[H]                       
\begin{center}     
\epsfig{figure=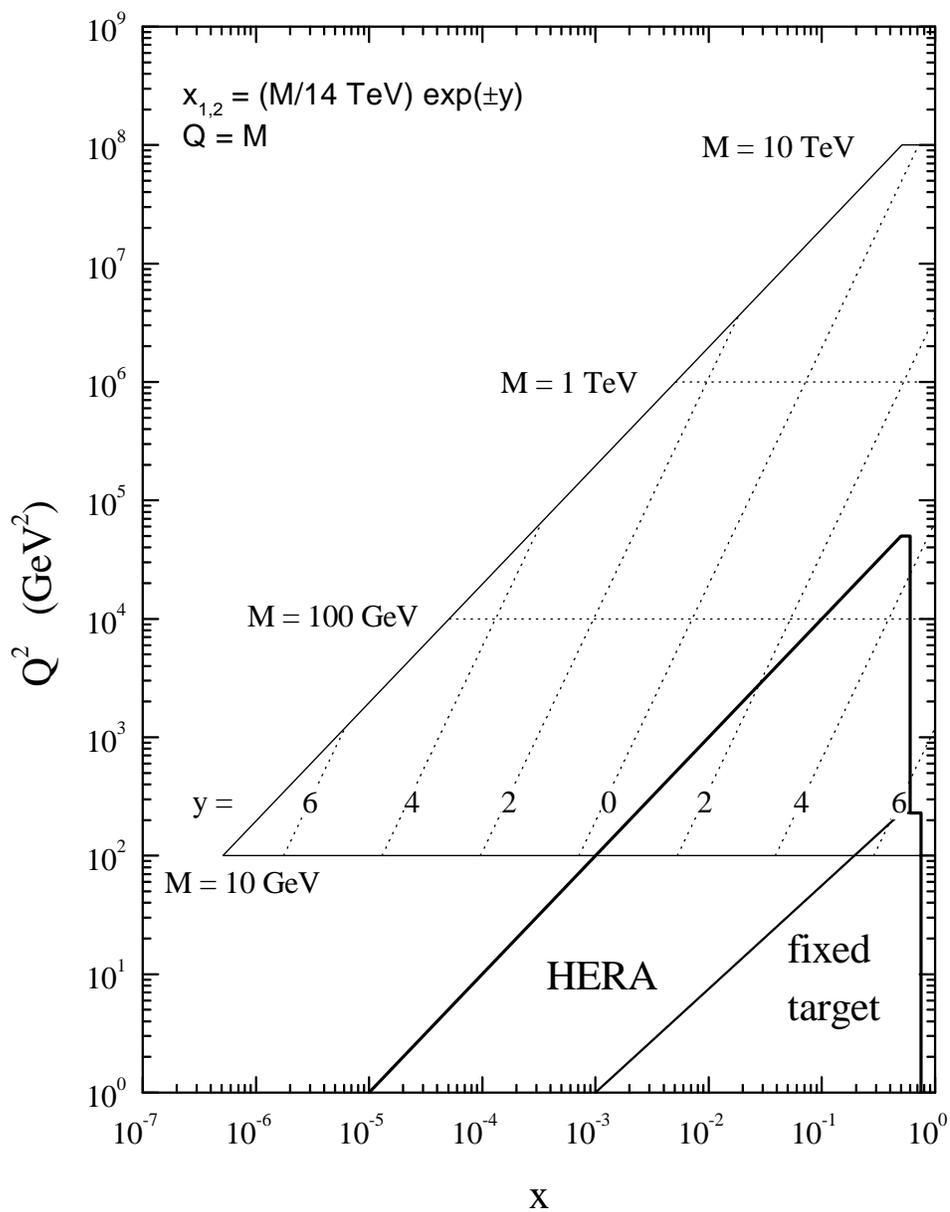,height=20cm}
\end{center}    
\vspace{-1cm}
\caption[]{Values of $x$ and $Q^2$ probed in the production of an object
of mass $M$ and rapidity $y$ at the LHC, $\sqrt{s} = 14\; \TeV$.}
\label{fig:LHCpartons}
\end{figure}

Before discussing  the $W$ and $Z$ cross sections in detail, we begin
with some general remarks about the
range of $x$ and $Q^2$  values that  are probed by
LHC cross sections. Figure~\ref{fig:LHCpartons} shows
the values corresponding to
the production of a heavy object (e.g. a $W$ or Higgs boson, a $ t\bar t$
pair, a multijet final state etc.) of mass $M$ and rapidity $y$.
We assume leading order kinematics, so that $x = M\exp(\pm y)/\sqrt{s}$
and $Q=M$. As an example, a $W$ boson ($M = 80\; \GeV$)
produced at rapidity $y=3$ corresponds to the annihilation of quarks
with $x=0.00028$ and $0.11$, probed at $Q^2 = 6400\; \GeV^2$.
Notice that quarks with these $x$ values are already `measured' in deep
inelastic scattering (at HERA and in fixed--target experiments
respectively), but at much lower $Q^2$, see Fig.~\ref{fig:LHCpartons}.

\bigskip

\noindent {\large \bf 2. Parton sets used in this study}   
\medskip 

In Ref.~\cite{MRST} we presented a set of pdfs, labelled MRST, that
gave the best overall fit to the global set of data in that analysis.
Associated with this set was a `central' gluon distribution (at the starting
scale of $Q_0^2 = 1$~GeV$^2$) and a `central' value of $\as(M_Z^2)=0.1175$. This
set resulted from a NLO analysis in the $\msb$ scheme and we can regard
this particular set as a base from which we wish to explore the
degree of latitude that relevant parameters can range over
while still maintaining an acceptable description
of all the data.   Already two degrees of freedom were explored in 
Ref.~\cite{MRST}, (i) the variation of the average transverse momentum
$\langle k_T \rangle$ in prompt photon production, which  resulted in a 
range of gluon distributions at large $x$, and (ii) the variation
in the value of $\as(M_Z^2)$. These uncertainties were reflected in 
alternative sets of pdfs MRST($\gup$), MRST($\gdown$), MRST($\asup$) and
MRST($\asdown$) in which the extremes of an allowed range of values
are taken and still yield acceptable fits. Again these sets were relevant to
a NLO analysis in the $\msb$ scheme.\footnote{In ref.~\cite{MRST2} we 
extended the analysis to provide equivalent sets of pdfs in the DIS scheme
and also at LO.} 

As we shall see below, the level of precision with which we should be able
to make predictions for $W$ and $Z$ cross sections at the Tevatron and LHC
is of the order of $\pm 5\%$. It goes without saying that the 
numerical precision with which the pdfs are extracted and evolved should 
be much smaller than this.  It is therefore worrying that a
comparison of the $\sigma_W$ values obtained using the default MRST partons
and the recent (default)  CTEQ5 partons \cite{CTEQ5} reveals differences at the level
of $5\%$, even though the data sets, theoretical assumptions
 and fitting procedures are largely the same in the two analyses. This difference
 has prompted us to re-examine the numerical accuracy of the MRST evolution
 code. As a result we discovered two small errors which affect the resulting
 pdfs at the $\sim 1\%$ level. Further details can be found in the Appendix.
The basic five parton sets used   in this analysis are {\it not} therefore
the five sets presented in Ref.~\cite{MRST}, but rather their corrected versions,
which we denote by MRST99, MRST99($\gup$), 
\ldots etc.\footnote{The {\tt FORTRAN} code for all the MRST99 parton sets
described in this paper  can be obtained from
{\tt http://durpdg.dur.ac.uk/HEPDATA/PDF},  or by contacting 
{\tt  W.J.Stirling@durham.ac.uk}.}
The $W$ and $Z$ Tevatron total cross section predictions
 using the MRST99 sets are some $2\%$
larger than the corresponding MRST values presented in 
Ref.~\cite{MRST}.\footnote{In 
contrast, the LHC cross sections are essentially unchanged, due to a slight
$x$ dependence of the correction factor.} The difference between the CTEQ5 and MRST
cross sections is only slightly reduced by using the corrected MRST99 partons. We will
return to this below.

Varying the  gluon and $\as$ about their central values does not span the full
range of uncertainties in the pdfs.
Obviously, varying the normalisation of sets of input data to the analysis
will result in a corresponding variation of the output pdfs and so this is
an additional degree of uncertainty which we now explore. Since in this
work we are primarily interested in uncertainties which most affect 
$W$ and $Z$ production at collider energies, i.e. uncertainties at small $x$,
we focus mainly on the 
normalisations of the HERA structure function data. The quoted normalisation uncertainties
of $F_2$ from H1 and ZEUS vary with $Q^2$, but a mean value of 
2.5\%~\cite{H1,ZEUS} is appropriate. We then proceed to make two fits
to all the data where we renormalise the HERA data first upwards
by 1.025, and then downwards by 0.975, and
allow the normalisations of the remaining DIS datasets to vary within
their quoted uncertainties in order to get the best fit. We label these two
pdf sets as MRST($\qup$) and MRST($\qdown$).  

\begin{figure}[H]                       
\begin{center}     
\epsfig{figure=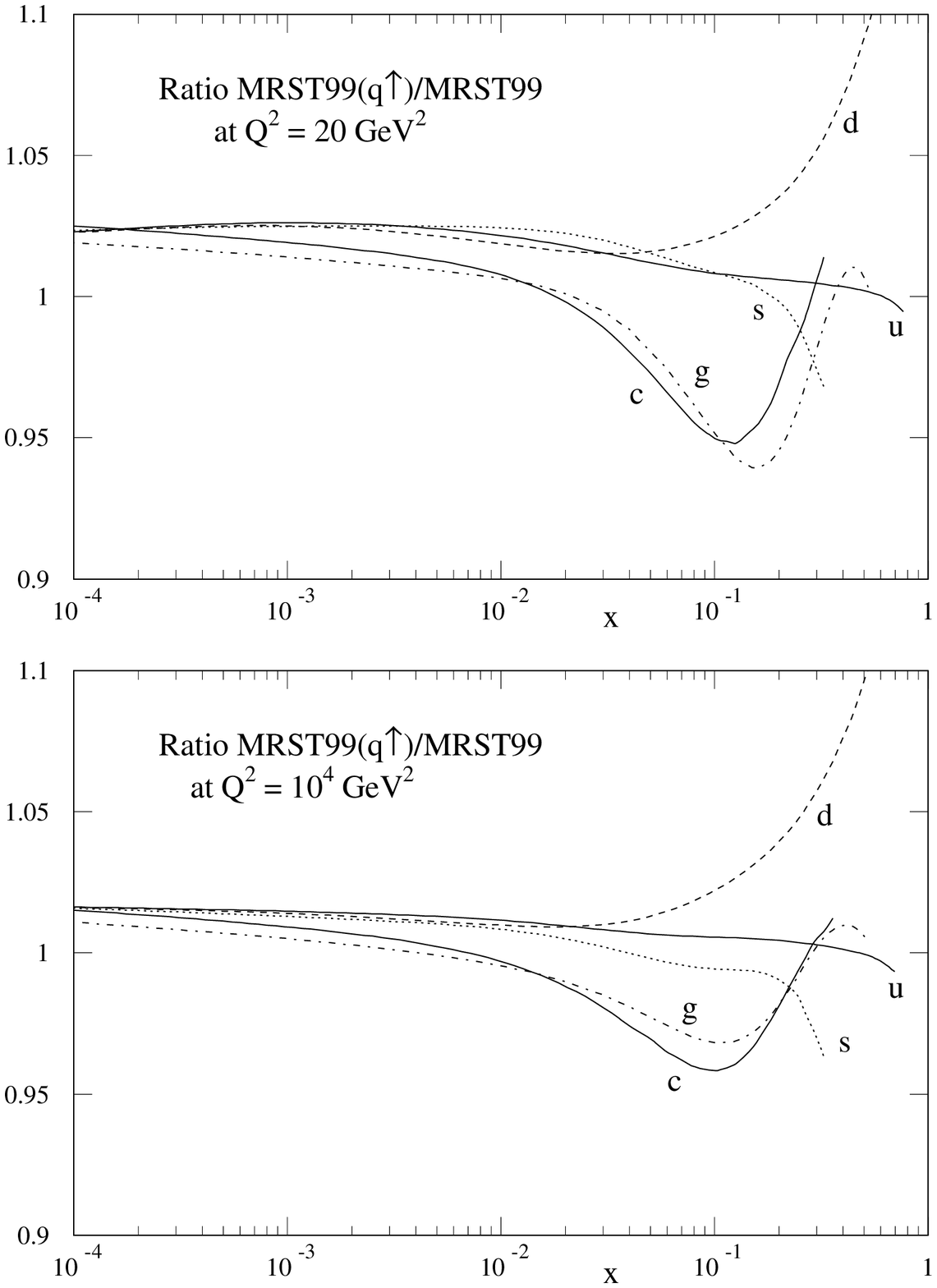,height=20cm}
\end{center}    
\vspace{-1cm}
\caption[]{Comparison of the MRST($\qup$) and default MRST pdfs at
$Q^2 = 10$ and 10$^4$~GeV$^2$.}
\label{fig:ratin}
\end{figure}

\begin{figure}[H]                       
\begin{center}     
\epsfig{figure=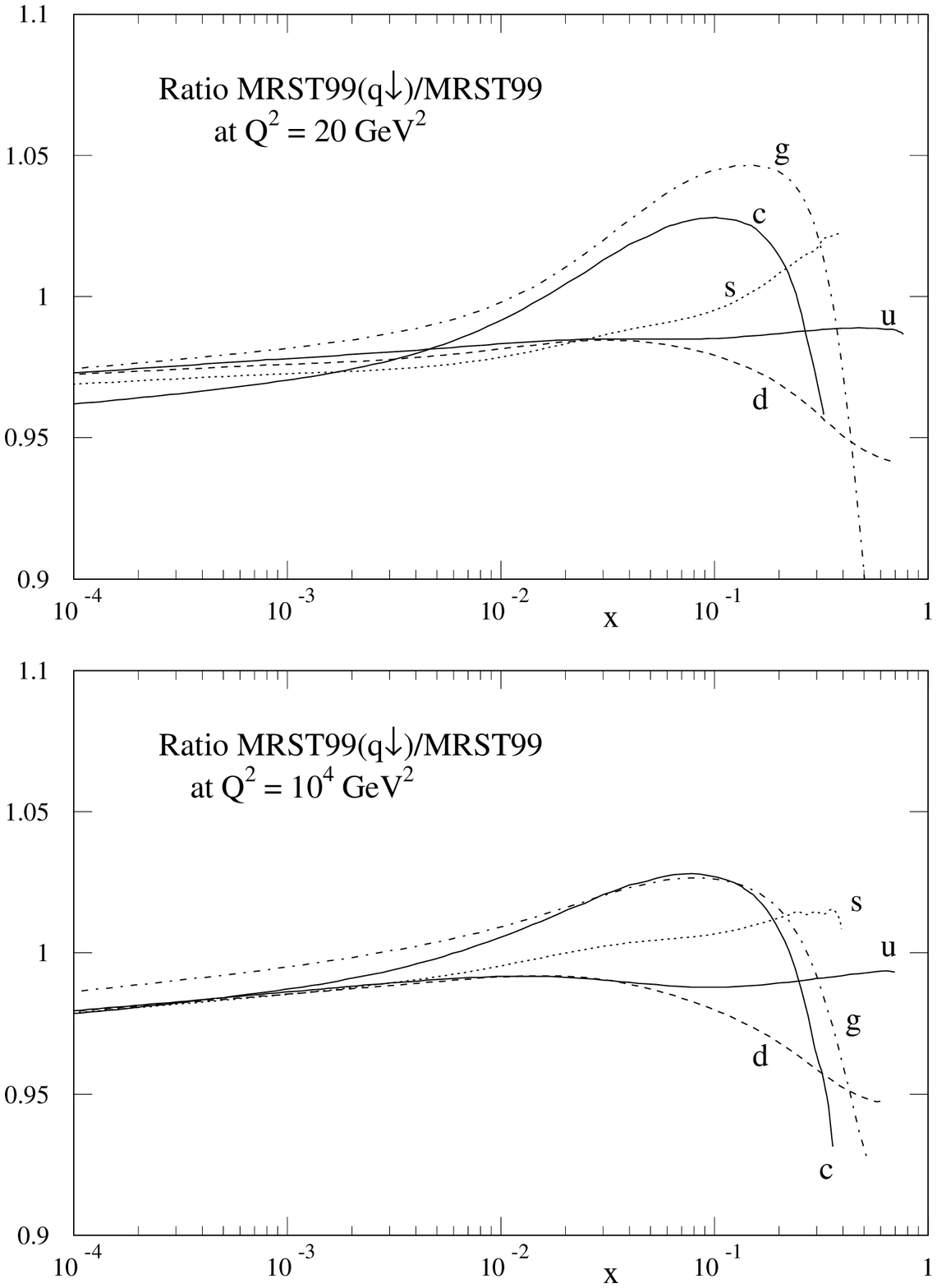,height=20cm}
\end{center}    
\vspace{-1cm}
\caption[]{Comparison of the MRST($\qdown$) and default MRST pdfs at
$Q^2 = 10$ and 10$^4$~GeV$^2$.}
\label{fig:ratin1}
\end{figure}

In Figs.~\ref{fig:ratin} and~\ref{fig:ratin1} we show comparisons of these
two pdf sets with the default set at two values of $Q^2$, a low $Q^2$ value 
typical of the $Q^2$ range of the fitted DIS data, and a high $Q^2$ value typical
of that probed in $W$, $Z$ production at hadron colliders. One can see that
when $u$ and $d$ increase to describe the larger $F_2$, momentum conservation
demands a decrease in the gluon distribution and vice versa.  
Because of the way charm
and bottom pdfs are generated (by photon--gluon fusion), the heavy flavour pdfs
simply `mirror' the modified gluon. Increasing the normalisation also
produces, at large $x$, a slightly increased $d/u$ ratio (and vice versa)
but this corresponds to  a very small variation of $F_2^n/F_2^p$. For example,
 at $x=0.675$ and $Q^2 = 35$~GeV$^2$, $F_2^n/F_2^p = 0.447\pm0.008$.
  
\begin{figure}[H]                       
\begin{center}     
\epsfig{figure=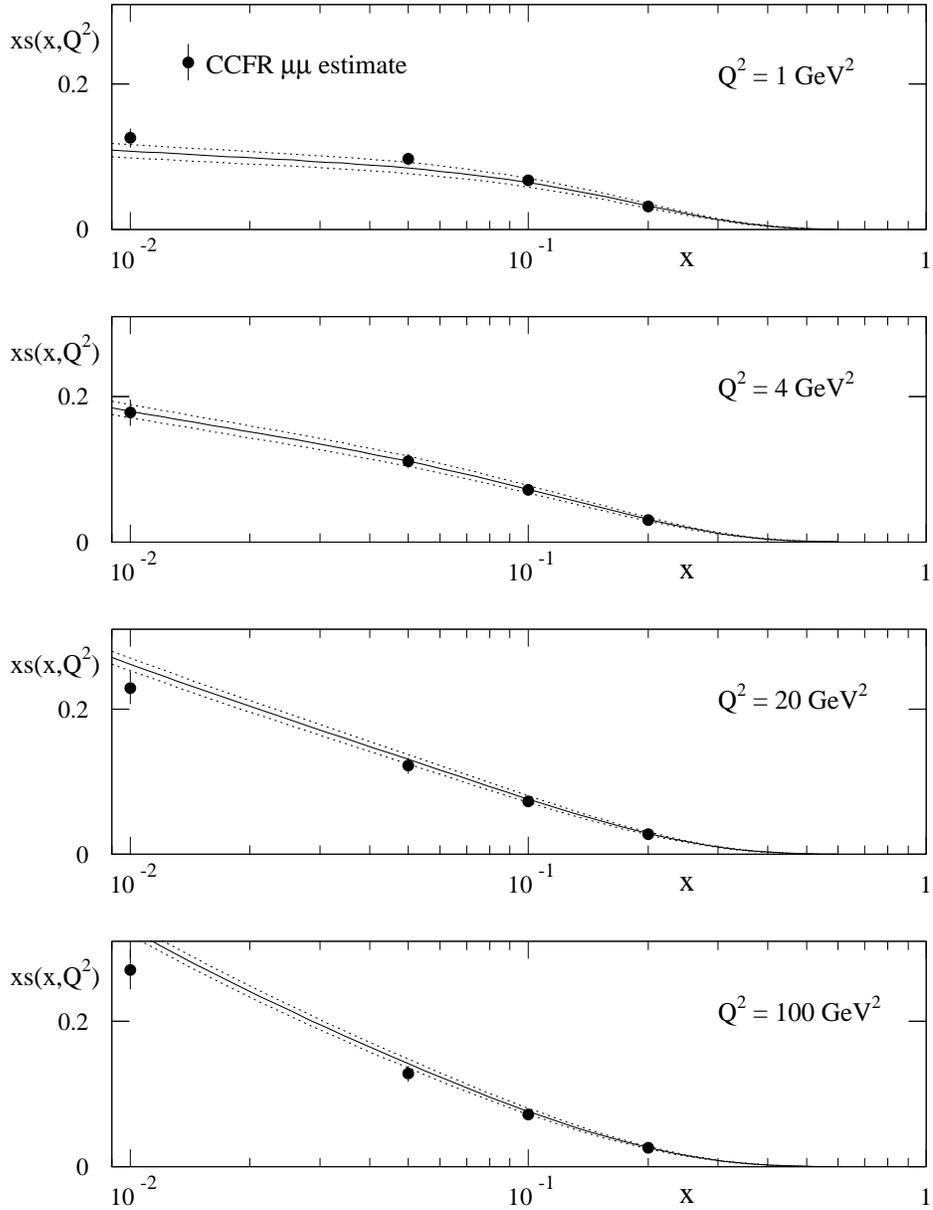,height=18cm}
\end{center}    
\vspace{-1cm}
\caption[]{Comparison of the strange sea from the NLO analysis of
CCFR~\cite{CCFRMM} dimuon production with the default MRST (solid line) 
and a variation of $\pm 10\%$ in the proportion (strange sea)/(total sea)
at $Q_0^2 = 1$~GeV$^2$ representing the input to the pdf sets MRST($\sup$)
and MRST($\sdown$).}   
\label{fig:strange}
\end{figure}

This variation of the normalisation effectively implies that the pdf sets
MRST($\qup$) and MRST($\qdown$) give a measure of overall uncertainty on the
$u$ and $d$ quarks. Next we try to get a corresponding measure of uncertainty on the
strange quark distribution. The strange sea is assumed to be identical in
shape to the non-strange sea at the starting scale  $Q_0^2$, but there is a 
strong flavour asymmetry due presumably to the larger mass of the strange quark. 
A suppression factor 
$2 \bar s/(\bar u + \bar d)$, at $Q_0^2 = 1$~GeV$^2$ of 0.5
is consistent with the di-muon production data of 
CCFR~\cite{CCFRMM}. Fig.~\ref{fig:strange} shows that a $\pm 10 \%$ variation  
of this suppression is a realistic measure of the uncertainty in the 
CCFR estimate. Consequently we provide two more sets of pdfs corresponding
to these upper and lower estimates for the strange sea suppression,
labelled MRST($\sup$) and MRST($\sdown$)  

The only parameter characterising the charm distribution in the MRST analysis 
is the charm mass,
$m_c$. In the updated default set $m_c = 1.43$~GeV and we find a satisfactory
description of $F_2^c$ provided the variation of $m_c$ is within 0.15~GeV~\cite{RT}.
So to complete our family of pdfs which explore individual degrees of
uncertainty, we provide two sets MRST($\cup$) and MRST($\cdown$) in which
$m_c$ = 1.28 and 1.58~GeV respectively. A summary
table of this `MRST family' is given in Table~1. Note that
an additional set MRST($\du$)) will be introduced and described in
Section~4.

\begin{table}[htb]
\begin{center}
\begin{tabular}{|l|l|l|}\hline
set & $\as(M_Z^2)$ & comment \\ \hline
  MRST   &  0.1175 & default set \\
MRST($\gup$)  &  0.1175 & larger gluon at large $x$ \\
MRST($\gdown$)   &  0.1175 & smaller gluon at large $x$ \\
MRST($\asup$)   &  0.1225 & larger $\as$ \\
MRST($\asdown$)  &  0.1125 & smaller $\as$ \\
MRST($\qup$)   &  0.1178 & larger ($\sim +2.5\%$) quarks at large $x$ \\
MRST($\qdown$)  &  0.1171 & smaller ($\sim -2.5\%$) quarks at large $x$  \\
MRST($\sup$)   &  0.1175 & larger ($\sim +10\%$)  strange quarks  \\
MRST($\sdown$)  &  0.1175 & smaller ($\sim -10\%$) strange quarks  \\
MRST($\cup$)   &  0.1175 & larger charm quarks ($m_c = 1.28$~GeV)  \\
MRST($\cdown$)  &  0.1175 & smaller charm quarks ($m_c = 1.58$~GeV) \\
MRST($\du$)  &  0.1175 & larger $d/u$ ratio at large $x$ \\
\hline
\end{tabular}
\label{tab1}
\caption{The various MRST99 pdfs used in the present study (see the Appendix). 
Note that the statements 
about the variations of the different quarks and gluons are relative to the 
default MRST99 set and apply at or just above the starting scale 
$Q_0^2 = 1\; \GeV^2$.}
\end{center}
\end{table}

\bigskip
                                     
\noindent {\large \bf 3. $W$ and $Z$ total cross sections}                                                                                        
\medskip 

The total $W$ and $Z$ cross sections probe quark distributions in a fairly narrow
$x$ region around $x \sim M_V/\sqrt{s}$ and at 
$Q^2 \sim M_V^2 \sim 10^4\; \GeV^2$.\footnote{For the subset of 
$W$s and $Z$s produced with large
transverse momentum the next-to-leading order 
$qg$ scattering processes are also important, and as a result  there is 
additional sensitivity to the  gluon distribution in roughly the same $x$ region -- 
we will not discuss this further here.}
The inclusive rapidity distributions $d\sigma/dy_V$ select particular
values of parton $x$ (see Fig.~\ref{fig:LHCpartons}) but, at least in the case
of $W^\pm$ production, are presumably more difficult to measure. They will be discussed
in the following section.

We have already argued that the dominant theoretical uncertainty comes from 
parton distributions, and so we fix all electroweak parameters at 
standard values \footnote{The Standard Model values  $B(W \to l \nu)  = 
0.1084$ and $B(Z\to l^+l^-) = 0.03364$ are used. The electroweak 
boson masses are taken to be $M_W = 80.43\; \GeV$ and 
$M_Z = 91.1887\; \GeV$.} and vary only the input pdfs (and $\as$ where appropriate).

\begin{figure}[H]
\begin{center}     
\epsfig{figure=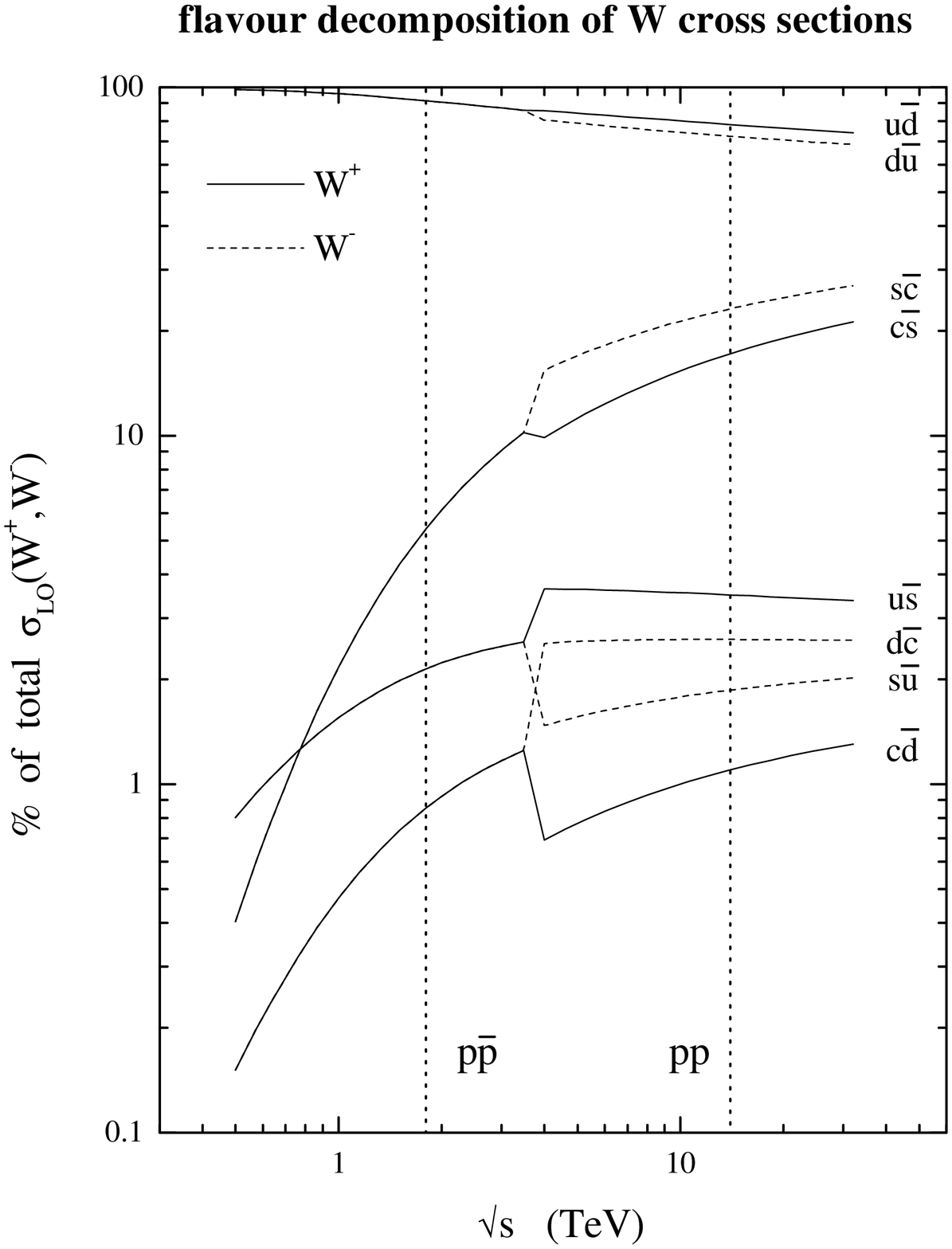,height=20cm}
\end{center}    
\vspace{-2cm}                                                                
\caption{Parton decomposition of the $W^+$ (solid line) and $W^-$
(dashed line) total cross sections in $\pp$ and $pp$ collisions.
Individual contributions are shown as a percentage of the total
cross section in each case. In $\pp$ collisions the decomposition is the
same for $W^+$ and $W^-$.}
\label{fig:LHCwdec}
\end{figure}

\begin{figure}[H]
\begin{center}     
\epsfig{figure=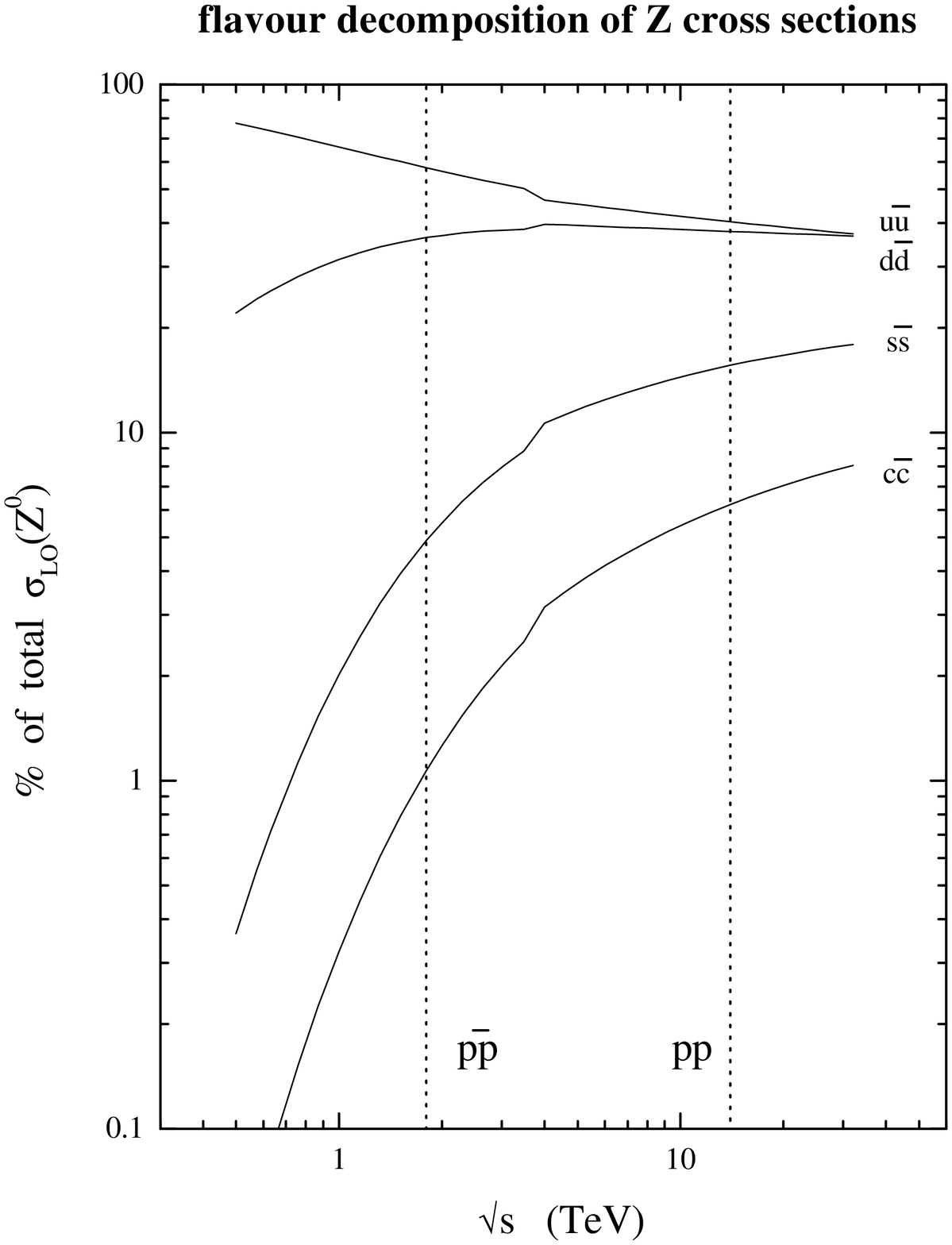,height=20cm}
\end{center}    
\vspace{-2cm}                                                                
\caption{Parton decomposition of the $Z^0$
total cross sections in $\pp$ and $pp$ collisions.
Individual contributions are shown as a percentage of the total
cross section.}
\label{fig:LHCzdec}
\end{figure}

First, we consider the parton flavour decomposition of $\sigma_{W^+}$,  $\sigma_{W^-}$
and  $\sigma_{Z}$. Since the cross sections are dominated by the 
leading-order $q \bar q \to V$ processes, it is sufficient to 
consider the relative contributions of these to the total.
Figures~\ref{fig:LHCwdec} and \ref{fig:LHCzdec} show  how the 
relative contributions of the various $q\bar q$ processes change 
with collider energy. We split the collider energy range
at $\sqrt{s} = 4\; \TeV$, and assume proton--antiproton and proton--proton collisions
below and above this value. For $W^\pm$ production (Fig.~\ref{fig:LHCwdec})
we see that the $u\bar{d}\to W^+$ and $d\bar{u}\to W^-$ contributions dominate at
all collider energies. These are dominated
by  valence--valence and valence--sea scattering
at $\pp$ and $pp$ colliders respectively.
The next largest contributions come from 
$c\bar{s}\to W^+$ and $s\bar{c}\to W^-$. Although these are sea--sea processes, they dominate
the Cabibbo suppressed $u\bar{s} \to W^+$ etc. valence--sea contributions.
Perhaps the most significant result from Fig.~\ref{fig:LHCwdec}
 is that  strange--charm
scattering is a relatively large component of the total cross sections at LHC ---
approximately $17\%$ for $W^+$ and $23\%$ for $W^-$. This contrasts with 
their $5\%$
contribution at the Tevatron. The remaining scattering processes contribute
between $1\%$ and $3\%$ at LHC.

The corresponding situation for $Z^0$ production is somewhat 
simpler, see Fig.~\ref{fig:LHCzdec}. At LHC
energies we have $ u\bar{u} \sim d\bar{d} \gg s\bar{s} \gg c\bar{c}$, in line
with the ordering of the parton distributions at the relevant $x$ and $Q^2$ 
values.\footnote{Note that the weak neutral couplings are slightly different for
$u$-- and $d$--type quarks, and are taken into account in the curves
shown in Fig.~\ref{fig:LHCzdec}.}
At the Tevatron, the  $s\bar{s}$ and $c\bar{c}$ contributions are relatively smaller,
and the $u\bar{u}$ contribution is correspondingly larger.

We have already shown in Ref.~\cite{MRST} that the Standard Model
predictions for $\sigma(W)$ and $\sigma(Z)$ at the Tevatron ($\sqrt{s} = 1.8\; \TeV$)
using the default MRST98 pdf set 
are in excellent agreement with the CDF and D0 measurements. Here we repeat the comparison 
now including the extended set of MRST99 
pdfs described in the previous section (and the Appendix). 
From Fig.~\ref{fig:LHCwdec},
we would not expect any significant effect from varying the charm and strange partons, and indeed
the $W$ and $Z$ cross sections calculated with the $\sup,\sdown,\cup,\cdown$ sets agree
with the default MRST99 cross section to within less than one percent.
On the other hand, the effect of
changing  the overall quark normalisation should be  directly reflected in the cross sections.
Figs.~\ref{fig:TEVwzdat} shows  the predictions of the first seven MRST99 sets 
(the default MRST99 set, together with  six other  $X\updownarrow$ sets ($X=g,\as,q$) 
listed in Table~1) for $\sigma_W$ and $\sigma_Z$, together 
with the experimental data, all normalised to 
$\sigma({\rm MRST})$.\footnote{Note that an important component of the experimental
error is due to  the luminosity measurement and uncertainty. The 
latter is quoted as $\pm 3.6\%$ for CDF and $\pm 5.4\%$ for D0 
\cite{CDFSIGW,D0SIGW}.  In addition, the value assumed for 
the total $p \bar p$ cross section is slightly different for the two 
experiments, and this may account in part for the systematically smaller
D0 cross sections displayed in Fig.~\ref{fig:TEVwzdat}.}
There is an overall spread of approximately $\pm 3\%$ about the default 
prediction,\footnote{A measure of the  scale dependence of these predictions is  
obtained by using instead $\mu = M_V/2$ and $2 M_V$. The effect is shown 
as error bars on the default prediction and is evidently very small.} 
significantly smaller than the current experimental errors. 
The variations in the predictions are easily understood. At these (small) 
$x$ values, the quark distributions {\it increase} with increasing $\mu^2$. 
The larger the $\alpha_S$ the faster the increase, and so   
$\sigma_V(\asdown) < \sigma_V({\rm MRST}) < \sigma_V(\asup)$.                                                     
The different gluon distributions also give rise to differences 
in the $\sigma_V$ predictions.  In this $x \sim 0.05$ region, 
the ordering of the gluon  distributions is 
$\gup < g({\rm MRST}) < \gdown$, see Fig.~2 of Ref.~\cite{MRST}
The larger the gluon the more rapid the DGLAP evolution, 
and so  $\sigma_V(\gup) < \sigma_V({\rm MRST}) < \sigma_V(\gdown)$. 
The gluon variation is slightly smaller than the $\alpha_S$ variation.
The largest variation (approximately $\pm 3\%$) comes from the overall
quark normalisation. Naively, one might have expected a variation
twice as large, reflecting the variation in the $\qup\downarrow$  
$u$ and $d$ partons near the starting scale, but the effect of
DGLAP evolution to $Q^2 \sim 10^4\; \GeV^2$ is to suppress the differences.
                                                                                                      
We may conclude from Fig.~\ref{fig:TEVwzdat} that 
the net uncertainty in the 
$\sigma_{W,Z}$ predictions at the Tevatron is no more than 
about $\pm 4\%$ and is largely due to the normalisation
uncertainty in the input $u$ and $d$ distributions, although
DGLAP evolution dilutes this effect by about a factor of 2.

Figure~\ref{fig:LHCwzdat} shows the corresponding total cross
sections at the LHC $pp$ collider. We have now also included the predictions
from the $\sup\downarrow, \cup\downarrow$ sets, but again we see that
the effect on the total cross section is minimal. Even though the
strange--charm contribution is relatively more important, DGLAP evolution
has reduced the variation in the partons themselves. Interestingly the
variations in the cross sections due to the gluon and the quark normalisation
are approximately the same as at the Tevatron, but now the largest variation
comes from $\as\updownarrow$. the reason for this is that DGLAP evolution
is more rapid at the smaller $x$ values probed at the LHC: schematically
we have $\partial q / \partial \log Q^2 \sim \as C$ where the (positive at small $x$)
 coefficient $C$ increases as $x$ decreases. Therefore the effect on the evolved $q(x,Q^2)$
 of varying $\as$ is magnified as $x$ decreases. Bearing in mind that our $\as$ variation
 is somewhat conservative (the `world average' error is now regularly quoted as $\pm 0.004$
 or even less), it would appear from Fig.~\ref{fig:LHCwzdat} that $\pm 5\%$ is a reasonable
 upper limit on the pdf uncertainty on $\sigma_{W,Z}$ at LHC.
 
In Fig.~\ref{fig:LHCwzdat} the (unequal)  $W^+$ and $W^-$ 
cross sections have been added together. It is interesting to consider what can be learned from
the {\it difference} between these, in particular from the ratio $R_\mp = \sigma_{W^-}/\sigma_{W^+}$.
Ignoring the contributions from strange--charm scattering (which contribute equally
to  $\sigma_{W^+}$ and  $\sigma_{W^-}$), we have
\beq
R_\mp  \approx { d \bar{u}  \over u \bar{d} } = {d \over u} \cdot {\bar{u} \over \bar{d}} \; .
\eeq
Notice that in the region of $x$ we are considering, {\it both} fractions on the
right-hand side are $< 1$. This raises the question of whether we can obtain any information
on the ratio $\bar u / \bar d$, assuming that $d/u$ is already well pinned down.
Figure~\ref{fig:LHCwrat} shows the (proton--proton) collider energy dependence
of the cross section ratio calculated using MRST (solid curve).
 For comparison we also show (dashed curve) the predictions for a
modified version of MRST in which  $\bar u $ is set equal 
to  $\bar d$.\footnote{Formally, we replace the $\bar u$ and $\bar d$ pdfs appearing
in the cross section expressions by their average.}
Unfortunately the difference is small, and decreases with increasing $\sqrt{s}$.  
The reason for this is that in the MRST fits, it is {\it assumed} that the asymmetry
$(\bar d -\bar u)/(\bar d + \bar u)$ goes rapidly to zero as $x \to 0$. This trend is
certainly suggested by the E866 Drell-Yan asymmetry data \cite{E866}. However these
data extend down only to $x \sim 0.05$, well above the $x$ region probed
by $\sigma_W$ at LHC.  In principle one could investigate the effect of 
`pathological' behaviour of  $\bar d - \bar u$ as $x \to 0$ on $R_\mp$.

At $\sqrt{s} = 14$~TeV, the predictions for  $R_\mp$ using all twelve
MRST sets listed in Table~1  lie in the range
\beq
R_\mp =  0.731 \pm 0.005
\eeq
with the central value corresponding to the MRST prediction. Evolution, normalisation
etc. effects all tend to cancel in the ratio, leaving this as a `gold-plated' 
prediction of the Standard Model. 

Before considering the $W$ and $Z$ rapidity distributions at the LHC, we
consider the total cross section predictions using the latest CTEQ5
parton distributions \cite{CTEQ5}. Since these are obtained in a similar way 
to the MRST partons, we would expect the (default, at least) 
predictions to lie well within the MRST99$\pm 5\%$ band. Surprisingly, 
this is not the case. Figures~\ref{fig:TEVwzctq} and \ref{fig:LHCwzctq} 
show the CTEQ5M and 
CTEQ5HQ\footnote{The 5M and 5HQ sets differ in their treatment
of the heavy quark distributions \cite{CTEQ5}, with the 5HQ distributions
more comparable to the MRST treatment \cite{MRST}. In order to maintain
the same charm contribution to the $F_2$ structure function at small $x$, the 
5M and 5HQ gluon distributions are slightly different in this region,
and this leads to corresponding
differences  in the $W,Z$ cross sections, {\it cf.} the MRST
$g\updownarrow$ variation.} 
predictions compared to those of the various MRST99 sets. The CTEQ5M
predictions are about 4\% (7\%) larger at the Tevatron (LHC) than those
of MRST99!  The CTEQ5HQ predictions are even larger, lying well
outside the $\pm 5\%$ band.

The differences between the MRST99 and CTEQ5 cross section predictions
are due to corresponding differences in the underlying quark distributions.
This is illustrated in Fig.~\ref{fig:t995hq}, which shows the ratio of CTEQ5HQ and MRST99
partons at two representative $Q^2$ values, $10$ and $10^4$~GeV$^2$.
The most striking difference is in the gluon at large $x$. The MRST and CTEQ5 
analyses put more emphasis in the global fit
on the high $p_T$ (fixed target) 
prompt photon and (Tevatron collider) jet data respectively, and this is directly
reflected in the resulting gluon pdfs. Since the sea quark pdfs are driven
by $ g \to q \bar q$ in the DGLAP evolution, the same behaviour is seen
in the ratio of
 the strange and charm distributions in Fig.~\ref{fig:t995hq}.\footnote{The smaller
 choice of $m_c = 1.3$~GeV in the CTEQ5 analysis, compared to $m_c = 1.43$~GeV
 for MRST99, also enhances the difference in the charm distributions.}
 Of more relevance in the present context are the differences in the $u$ and
 $d$ quarks for $x \lapproxeq 0.1$ at $Q^2 = 10^4$~GeV$^2$.
One might expect that the long evolution length would result in
essentially identical quark distributions,\footnote{Note that at the lower
$Q^2$ value in  Fig.~\ref{fig:t995hq} the impact of the different MRST
and CTEQ5 starting parameterisations and $Q^2$ cuts on the fitted DIS data 
can still be seen. At the higher $Q^2$ value
these differences have evolved away and the ratios are more uniform.} 
but the CTEQ5HQ $u$ and $d$ quarks are  between 2\%  (at $x \sim 0.1$)
and 5\% (at $x \sim 0.0001$)  larger than those of MRST99
at small $x$.  A possible reason 
for this discrepancy can be found in the numerical results
presented  in the Appendix.
                                                                                                      
\bigskip

\noindent {\large \bf 4. $W$ and $Z$ rapidity distributions}
\bigskip

The $W^+$ and $W^-$ rapidity distributions are symmetric about
$y_W=0$ and have a distinctively different shape at large $y_W$, see
Fig.~\ref{fig:LHCwraps}.\footnote{The curves are calculated using
NLO corrections only -- the NNLO corrections are not known at present.}
 The bump in the $W^+$ distribution
around $y_W =3$  
is caused by a very small $x$ $\bar d$ quark scattering off
a large $x$ valence $u$ quark. At large $y_W$, therefore, the 
rapidity distributions probe small $x$ (evolved) `HERA quarks'
scattering off large $x$ (evolved) `fixed target' quarks.
For $W^+$  and $W^-$ production, the latter are constrained 
by, for example, $F_2^{\mu p}$ and $F_2^{\mu n}$ structure
functions respectively.

The distributions themselves are presumably difficult to measure
accurately in practice --- it is the {\it lepton} rapidity distribution
that is measured instead. We will not discuss the experimental
considerations here, see for example \cite{DITTMAR} for more discussion.
However it is worth pointing out that the {\it ratio} of $W^\pm$ rapidity
distributions is sensitive to the ratio of $d$ and $u$ quarks. Thus
\bea
R_\mp(y_W) \equiv {d\sigma/dy_W(W^-) \over d\sigma/dy_W(W^+)}
  &\approx& { d(x_1) \bar{u}(x_2)  \over u(x_1) \bar{d}(x_2) } 
  =  {d(x_1) \over u(x_1)} \cdot {\bar{u}(x_2) \over \bar{d}(x_2)} \nonumber \\
& \approx &{d(x_1) \over u(x_1)}  \; , \qquad x_1 = 
\frac{M_W}{\sqrt{s}}\, \exp(y_W)  = 
(5.74 \times 10^{-3})\, \exp(y_W) \; ,
\label{eq:raprat}
\eea
where for simplicity we have ignored the contributions involving
 strange and heavier quarks.
Assuming that $\bar d / \bar u \to 1$ as $x \to 0$, 
the large-$y_W$ rapidity distribution ratio is therefore a direct measure of
the large-$x$ $d/u$ ratio of partons, at $Q^2 = M_W^2$. Interestingly, there
has been considerable recent interest in this ratio, since it is also
 measured by charged current scattering cross sections at HERA:
\beq
{d^2\sigma/dxdQ^2(e^+p \to \bar\nu X) \over 
d^2\sigma/dxdQ^2(e^-p \to \nu X)} \approx (1-y)^2 \; {d(x,Q^2) \over u(x,Q^2) } 
\eeq

Fig.~\ref{fig:LHCwraprat} shows the ratio $R_\mp(y_W)$ as a function of $y_W$. Note that only
the predictions of the MRST set are shown -- the other ten sets of
Table~1 give almost identical predictions. This is because the $\updownarrow$ 
variations quantified by these other sets largely cancel in the ratio.
Also shown (dotted line) is the $d/u$ ratio evaluated at 
$x = (5.74 \times 10^{-3})\, \exp(y_W)$ and $Q=M_W$. The convergence of the curves
at large $y_W$ is confirmation of the derivation in (\ref{eq:raprat}).
At central rapidities, the parton ratio falls below the $W$ cross section
ratio. This is due to the non-($u,d$) contributions and $\bar u \neq \bar d$
at the higher $x$ values probed. For example,
Fig.~\ref{fig:LHCwraps} shows the rapidity dependence 
of the strange--charm scattering contribution, which contributes
equally to $W^+$ and $W^-$ production. The $W$ bosons from this sea--sea
scattering process are confined to the central region.

As mentioned above, current information on
$d/u$ at large $x$ comes from fixed target $F_2^{\mu n}/F_2^{\mu p}$
measurements and the lepton asymmetry 
in $p \bar p \to W^\pm + X$, see for example \cite{MRST}. 
In the MRST fit,
NMC $n/p$ data are used to constrain the large-$x$ $d$--quark pdf in this way.
When the corresponding predictions for $\sigma_{CC}(e^+p)$ are compared
with the ZEUS data \cite{ZEUScc}, there is  some
evidence of a slight excess of data over theory in the largest $x$ 
($=0.42$) bin.
Could this imply that the $d/u$ ratio is being underestimated in the
standard global fits? Any attempt to increase $d/u$ at large $x$ 
in the global fit leads
to a direct conflict with the $n/p$ data. However, Bodek and Yang \cite{BODEK}
have recently stressed a point first made by Thomas and  Melnitchouk
(Ref.~\cite{THOMAS} and references therein)
that the $n/p$ data should be corrected for nuclear binding effects 
which, at large $x$, lead to a larger $d/u$ ratio, in `better' agreement
with the ZEUS data. This is an issue that has yet to be resolved,
and therefore independent information from the $W^\pm$ rapidity
ratio at LHC would be very valuable. 

To illustrate the likely size of the effect, we show in Fig.~\ref{fig:LHCwraprat}
the predictions of a modified version of MRST in which the $d/u$ ratio
is deliberately enhanced at large $x$ such as to mimic the size of
effect proposed in Ref.~\cite{BODEK}. The difference is only significant
for $y_W \gapproxeq 4$, which suggests that discrimination will
not be easy at the LHC.

Since it is the {\it charged lepton} rapidity distribution that is measured
in practice, we can see to what extent the differences between $W^+$ and $W^-$
production survive the leptonic decay process. Fig.~\ref{fig:LHClraprat}
shows the ratio $R^\ell_\mp$ of the charged lepton rapidity
distributions.\footnote{The lepton distributions are computed at leading order
with the full $V-A$ decay structure taken into account. The ratio is very
insensitive to higher-order effects.}
 A transverse momentum cut of $p_T^\ell > 25$~GeV$/c$ is imposed.
The solid line is again the MRST prediction, and the dashed line 
is the prediction of the enhanced $d/u$ set, as in Fig.~\ref{fig:LHCwraprat}.
Notice the qualitatively different behaviour of the ratio
at large rapidity. This is due to the $V-A$ structure of the 
$q\bar q' \to \ell \nu_\ell$ subprocess. Thus a $W^+$ produced
with large positive $y_W$ in $u \bar d \to W^+$, decays preferentially
to a $e^+$ with {\it smaller} rapidity $y_\ell$,  while a 
$W^-$ produced
with large positive $y_W$ in $d \bar u \to W^-$, decays preferentially
to a $e^-$ with {\it larger} rapidity $y_\ell$. The ratio
$R^\ell_\mp$ therefore {\it increases} as  $y_\ell \to y_{\rm max}$.
This is in direct analogy to the lepton asymmetry in $p \bar p \to WX$
changing sign and becoming negative at large $y_\ell$.
The effect of  enhancing the $d/u$ ratio at large $x$
is still just visible.

In the same spirit we can consider the ratio of $W^\pm$ and $Z^0$ rapidity
distributions. The analogue of (\ref{eq:raprat}) is now
\bea
R_{Z/W}(y) \equiv {d\sigma/dy(Z^0) \over d\sigma/dy(W^+) + d\sigma/dy(W^-)}
  \approx { \kappa_u\; u(x'_1) \bar{u}(x'_2) + \kappa_d\; d(x'_1) \bar{d}(x'_2) \over 
 \vert V_{ud}\vert^2 \left\{ u(x_1) \bar{d}(x_2) +  d(x_1) \bar{u}(x_2)  \right\}} 
\; .
\label{eq:zraprat}
\eea
Here the $u$ and $d$ contributions to the numerator are weighted by
the appropriate neutral current coupling factors:
\beq
\kappa_q = \cos^{-2}\theta_W\,  (v_q^2 + a_q^2) \approx 0.37 (0.48) \quad \mbox{for}\quad q=u(d) \; .
\eeq
Because $M_W \neq M_Z$, the pdfs in the numerator and the denominator
are no longer evaluated at exactly the same values of $x$. This means that
there is no simple analytic expression for the ratio  in the 
$y \to y_{\rm max}$ limit.
The prediction for $ R_{Z/W}(y) $ calculated using the MRST 
partons is shown in Fig.~\ref{fig:LHCwzraprat}. 
The prediction using the  MRST($d/u$-enhanced) set is essentially
the same --- there is little sensitivity
to the $d/u$ ratio. 
\bigskip

\noindent {\large \bf 5. Conclusions} 
\medskip

We have made a detailed study of the $W$ and $Z$ boson
total production cross sections and rapidity distributions at  the LHC 
proton-proton  collider. We have focused on the dependence of the cross sections
on the parton distributions --- the largest source of theoretical
uncertainty. We have shown that the two important effects are the 
DGLAP evolution uncertainty, due to the uncertainty in
$\alpha_S$ and to a lesser extent in the gluon, and the normalisation
uncertainty on the  DIS structure function data fitted in this analysis,
which is directly
reflected in a corresponding 
uncertainty in the parton distributions. Taking everything together,
we conservatively estimate an overall pdf uncertainty of approximately $\pm 5\%$
on the total cross sections at the LHC, only slightly larger than
the uncertainty on the corresponding Tevatron
cross sections.

We have also studied the rapidity distributions. If these can be accurately
measured, they allow the quark distributions to be probed over a wide
range of $x$. We showed, in particular, that the difference in the $W^-$
and $W^+$ distributions at large rapidity directly reflects the 
difference in the $d$ and $u$ distributions as $x \to 1$. A measurement
of the $d/u$ ratio in this way would be complementary to the 
information obtained from charged current $e^\pm p$ cross sections at
HERA. 

Finally, while comparing our total cross section predictions with
those based on the latest CTEQ5 distributions we noticed differences
that we believe reflect shortcomings in the accuracy of the 
evolution codes. We have corrected an error (at NLO) in the standard MRS
evolution code that brings it into exact agreement with the ``standard
evolution" benchmark program. The CTEQ5 $W$ and $Z$ cross sections are still
systematically larger than those using the corrected MRST partons, 
an effect we believe is due to a too rapid
CTEQ sea quark evolution.

\noindent {\large \bf Acknowledgements}  
    
This work was supported in part by the EU Fourth Framework
Programme `Training and Mobility of Researchers', Network
`Quantum Chromodynamics and the Deep Structure of
Elementary Particles', contract FMRX-CT98-0194 (DG 12 - MIHT).
\newpage 
                                                                       
\noindent {\large \bf Appendix} 
\bigskip
     
The parton sets used in this paper, MRST99, differ slightly from the original 
MRST \cite{MRST} distributions.  The main difference is due to the correction 
in the computer code of a factor $C_F \rightarrow C_A$ in one of the terms
appearing in the NLO $P_{gg}$ splitting function. The effect is to make the 
gluon, and hence the sea, evolve a little more rapidly at small $x$.  
We have also included a missing NLO $P_{cq}$ contribution which affects  
the charm evolution only.

It is informative to explain the present situation concerning evolution codes. 
The long evolution length to Tevatron and LHC energies, and the expected 
precision of the predictions of the weak boson production cross sections, 
can expose small defects in evolution codes.  
The unexpectedly large difference that we found between the cross sections 
predicted using MRST98 and CTEQ5 partons means that all codes should be 
carefully checked.  The difference appears to be due to the more rapid 
evolution of the sea quarks of CTEQ5.  A similar effect was found in a 
comparison of various NLO evolution codes made during the 
1996 HERA workshop \cite{WS96}.  Neither CTEQ, nor MRS, agreed completely with
the \lq\lq standard code\rq\rq~for a test evolution from $Q^2 = 4$ 
to 100~GeV$^2$.  In the interval $10^{-4} \lapproxeq x \lapproxeq 10^{-1}$, 
the CTEQ sea was some 4\% high, and the MRS sea was 1\% low and the gluon 
about 1.5\% low.  After including the corrections 
noted above, the MRST99 code is found to agree precisely with the \lq\lq 
standard\rq\rq~evolution, and consistent with exact momentum conservation 
as a function of $Q^2$, as it should be.  
The momentum fractions carried by the various partons as $Q^2$ increases 
are shown in Table 1.  Small violations of overall momentum conservation
can be seen for MRST98 and the CTEQ parton sets (CTEQ4M and CTEQ4HQ are
qualitatively similar). 

Since the MRST99 partons come from a new global fit to the data, a comparison 
with the MRST98 partons \cite{MRST} does not simply reflect the small 
correction to the evolution code.  
In particular, ZEUS 1995 SVX data \cite{ZSVX} is included in the MRST99 
analysis.  
Moreover we have increased the mass of the charm quark $m_c$ from 1.35 to 
1.43~GeV so as to maintain the description of the $F_2$ (charm) data. \\

\begin{table}[htb]
{\small
\begin{tabular}{|r|ccccccccc|}\hline
{\bf CTEQ5M} & $u_v$ & $d_v$ & $2u_{sea}$ & $2d_{sea}$ & $2s$ & $2c$ 
& $2b$ & $g$  & total \\ \hline 
$Q^2\; ({\rm GeV}^2) =$ 
  5 & 0.2843 & 0.1143 & 0.0559 & 0.0730 & 0.0374 & 0.0120 & 0.0000 
    & 0.4293 & 1.0063 \\
 10 & 0.2667 & 0.1072 & 0.0586 & 0.0746 & 0.0412 & 0.0174 & 0.0000 
    & 0.4424 & 1.0081 \\
 50 & 0.2374 & 0.0955 & 0.0631 & 0.0774 & 0.0477 & 0.0265 & 0.0059 
    & 0.4576 & 1.0110 \\
100 & 0.2278 & 0.0916 & 0.0646 & 0.0783 & 0.0498 & 0.0294 & 0.0097
    & 0.4607 & 1.0120 \\
500 & 0.2096 & 0.0843 & 0.0674 & 0.0800 & 0.0538 & 0.0350 
    & 0.0169 & 0.4667 & 1.0137 \\
1000 & 0.2031 & 0.0817 & 0.0684 & 0.0806 & 0.0552 & 0.0370 & 0.0194 
    & 0.4688 & 1.0142 \\
5000 & 0.1902 & 0.0765 & 0.0703 & 0.0818 & 0.0579 & 0.0409 
    & 0.0245 & 0.4727 & 1.0149 \\
10000 & 0.1855 & 0.0746 & 0.0710 & 0.0822 & 0.0589 & 0.0424 & 0.0263 
    & 0.4742 & 1.0150 \\ \hline\hline

 {\bf CTEQ5HQ} & $u_v$ & $d_v$ & $2u_{sea}$ & $2d_{sea}$ & $2s$ & $2c$ 
& $2b$ & $g$  & total \\ \hline 
$Q^2\; ({\rm GeV}^2) =$ 
  5 & 0.2858 & 0.1160 & 0.0579 & 0.0749 & 0.0379 & 0.0118 & 0.0000 
    & 0.4215 & 1.0058 \\
 10 & 0.2681 & 0.1088 & 0.0604 & 0.0763 & 0.0416 & 0.0171 & 0.0000 
    & 0.4354 & 1.0075 \\
  50 & 0.2386 & 0.0968 & 0.0646 & 0.0787 & 0.0479 & 0.0260 & 0.0058 
     & 0.4519 & 1.0104 \\
 100 & 0.2290 & 0.0929 & 0.0660 & 0.0796 & 0.0499 & 0.0290 & 0.0096 
     & 0.4554 & 1.0114 \\
 500 & 0.2107 & 0.0855 & 0.0686 & 0.0811 & 0.0538 & 0.0345 & 0.0167 
     & 0.4621 & 1.0130 \\
1000 & 0.2042 & 0.0828 & 0.0695 & 0.0816 & 0.0552 & 0.0365 & 0.0192 
     & 0.4645 & 1.0135 \\
5000 & 0.1912 & 0.0776 & 0.0713 & 0.0826 & 0.0579 & 0.0404 & 0.0242 
     & 0.4689 & 1.0142 \\
10000 & 0.1864 & 0.0756 & 0.0719 & 0.0830 & 0.0589 & 0.0418 & 0.0260 
     & 0.4705 & 1.0143 \\ \hline\hline

{\bf  MRST98} & $u_v$ & $d_v$ & $2u_{sea}$ & $2d_{sea}$ & $2s$ & $2c$ 
& $2b$ & $g$  & total \\ \hline 
$Q^2\; ({\rm GeV}^2) =$ 
  5 & 0.2799 & 0.1160 & 0.0599 & 0.0759 & 0.0414 & 0.0087 & 0.0000 
    & 0.4108 & 0.9926 \\
 10 & 0.2627 & 0.1089 & 0.0614 & 0.0764 & 0.0441 & 0.0130 & 0.0000 
    & 0.4245 & 0.9910 \\
 50 & 0.2341 & 0.0970 & 0.0641 & 0.0775 & 0.0487 & 0.0205 & 0.0062 
    & 0.4406 & 0.9887 \\
100 & 0.2240 & 0.0928 & 0.0652 & 0.0779 & 0.0504 & 0.0232 & 0.0096 
    & 0.4452 & 0.9882 \\
500 & 0.2061 & 0.0854 & 0.0670 & 0.0787 & 0.0533 & 0.0281 & 0.0156 
    & 0.4524 & 0.9866 \\
1000 & 0.1994 & 0.0826 & 0.0677 & 0.0790 & 0.0545 & 0.0300 & 0.0179 
    & 0.4551 & 0.9862 \\
5000 & 0.1868 & 0.0774 & 0.0690 & 0.0796 & 0.0567 & 0.0335 & 0.0222 
     & 0.4594 & 0.9846 \\
10000 & 0.1818 & 0.0753 & 0.0695 & 0.0799 & 0.0575 & 0.0349 & 0.0239 
     & 0.4609 & 0.9838 \\ \hline\hline

{\bf   MRST99}  & $u_v$ & $d_v$ & $2u_{sea}$ & $2d_{sea}$ & $2s$ & $2c$ 
& $2b$ & $g$  & total \\ \hline 
$Q^2\; ({\rm GeV}^2) =$ 
  5 & 0.2810 & 0.1136 & 0.0600 & 0.0760 & 0.0416 & 0.0073 & 0.0000 
    & 0.4197 & 0.9992 \\
 10 & 0.2640 & 0.1067 & 0.0616 & 0.0767 & 0.0444 & 0.0122 & 0.0000 
    & 0.4335 & 0.9991 \\
 50 & 0.2359 & 0.0953 & 0.0646 & 0.0780 & 0.0492 & 0.0205 & 0.0066 
    & 0.4482 & 0.9983 \\
100 & 0.2259 & 0.0913 & 0.0657 & 0.0786 & 0.0509 & 0.0235 & 0.0102 
    & 0.4521 & 0.9982 \\
500 & 0.2083 & 0.0842 & 0.0676 & 0.0795 & 0.0540 & 0.0287 & 0.0166 
    & 0.4584 & 0.9973 \\
1000 & 0.2016 & 0.0815 & 0.0684 & 0.0799 & 0.0552 & 0.0308 & 0.0190 
    & 0.4608 & 0.9971 \\
5000 & 0.1891 & 0.0764 & 0.0697 & 0.0805 & 0.0574 & 0.0345 & 0.0235 
     & 0.4648 & 0.9960 \\
10000 & 0.1841 & 0.0744 & 0.0703 & 0.0808 & 0.0583 & 0.0360 & 0.0253 
      & 0.4662 & 0.9953 \\ \hline
\end{tabular}
\caption{\label{tableone}Momentum fractions carried by quarks and gluon for the 
 CTEQ5M \protect\cite{CTEQ5}, CTEQ5HQ \protect\cite{CTEQ5}, 
 MRST98 \protect\cite{MRST} and the new 
 MRST99 parton sets, for $Q^2$ between 5 and
10$^4$ GeV$^2$. The fractions are obtained by numerically computing the
integrals $\int_{x_0}^1 xf_i(x,Q^2)dx $ where $x_0 = 10^{-5}$. 
The missing interval,
$0 < x < 10^{-5}$, accounts for the small departure of the MRST99 total
from 1.} 
}
\end{table}

\newpage
                            
\newpage    
         
\newpage    

\newpage

\newpage    
\newpage

\newpage

\begin{figure}[H]
\begin{center} 
\epsfig{figure=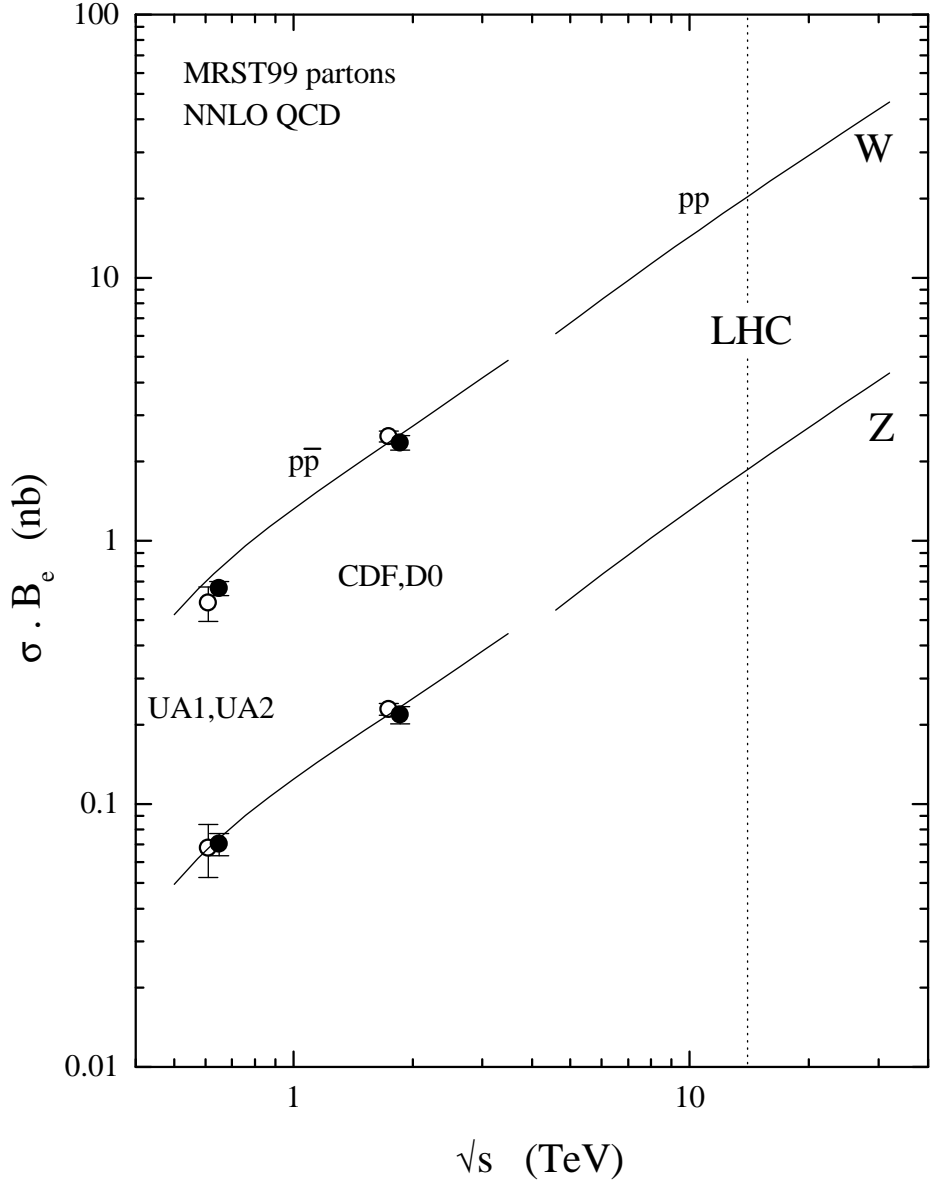,height=20cm}    
\end{center}    
\vspace{-1cm}
\caption[]{Predictions for the total $W, Z$ production
cross sections times leptonic branching
ratios in $\pp$ and $pp$ collisions,
as a function of the  collider energy $\sqrt s$. The default MRST99
partons are used.
Experimental measurements from     
UA1~\protect\cite{UA1SIGW}, UA2~\protect\cite{UA2SIGW}, 
CDF~\protect\cite{CDFSIGW}     
and D0~\protect\cite{D0SIGW} are also shown.}
\label{fig:WZsig}                            
\end{figure}                            
\newpage    
                            
\begin{figure}[H]                            
\begin{center}   
\epsfig{figure=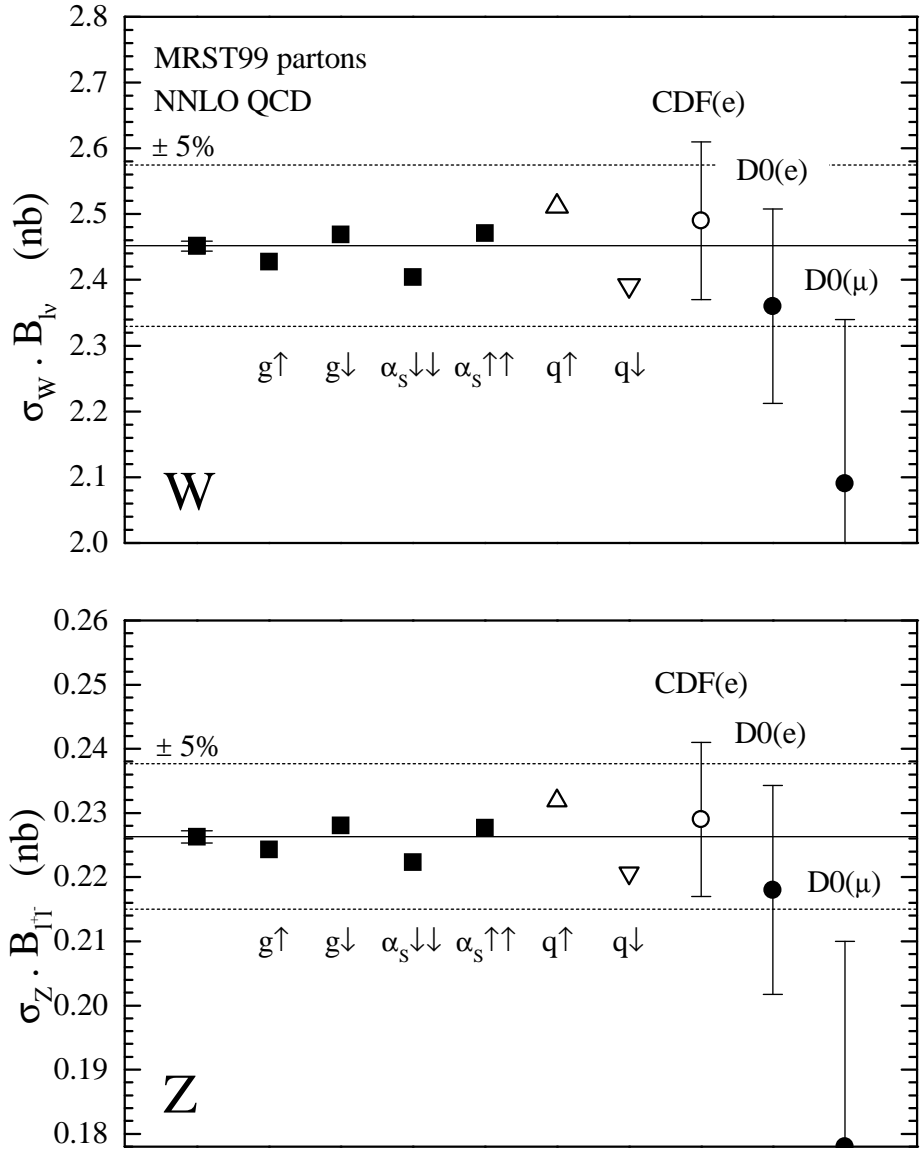,height=20cm}
\end{center}    
\vspace{-2cm}                                                                
\caption{Predictions for the $W$ and $Z$ production cross sections    
times leptonic branching ratios in     
$p \bar p$ collisions at  $1.8\; \TeV$
using the various MRST parton sets discussed in the text.
The error bars on the default MRST prediction
correspond to a scale variation of $\mu = M_V/2 \to 2 M_V$, $V=W,Z$.    
Experimental measurements from     
 CDF~\protect\cite{CDFSIGW} and D0~\protect\cite{D0SIGW} are shown.}
\label{fig:TEVwzdat}
\end{figure}                            
\newpage

\begin{figure}[H]
\begin{center}     
\epsfig{figure=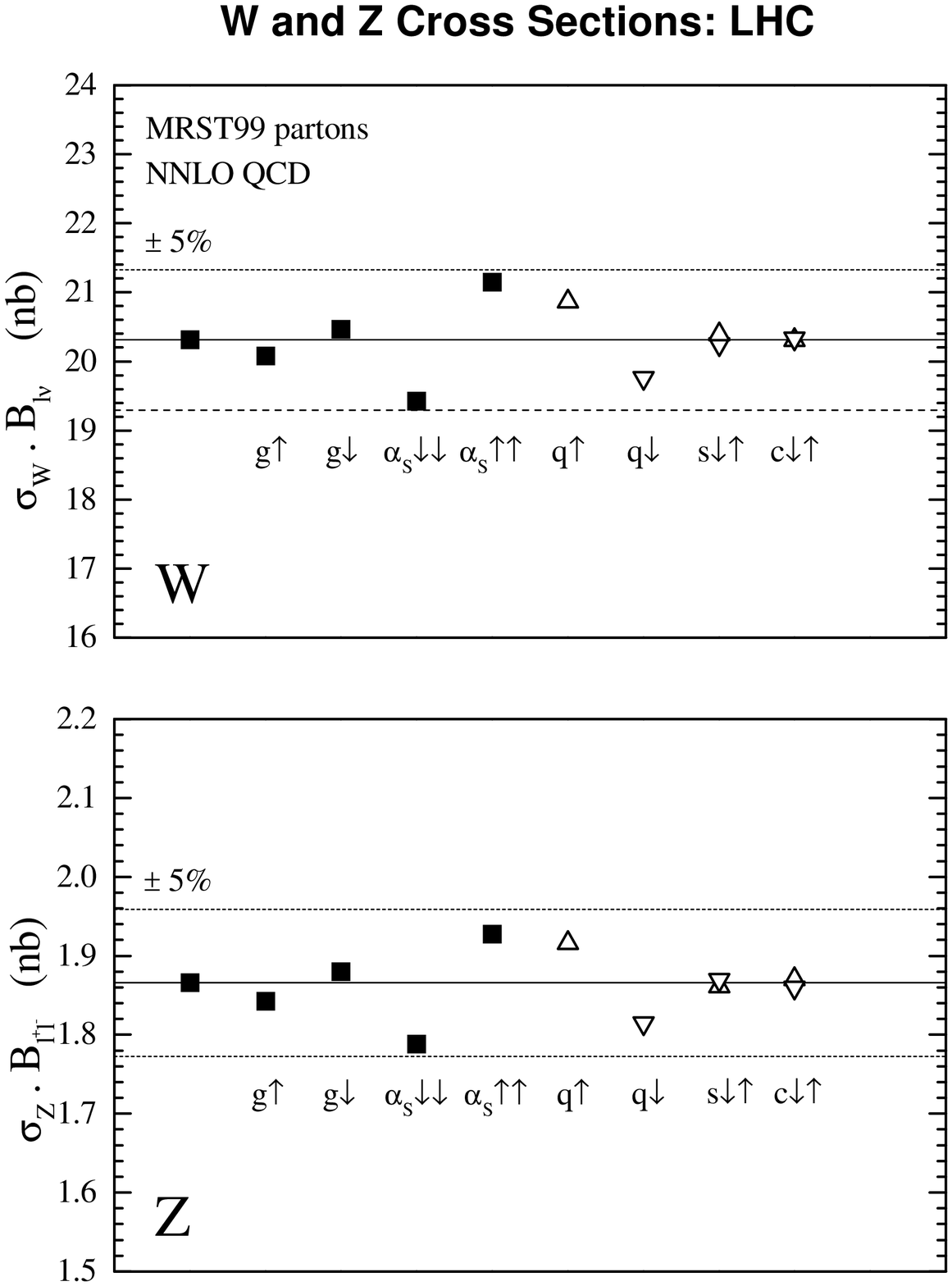,height=20cm}
\end{center}    
\vspace{-2cm}                                                                
\caption{Predictions for the $W$ and $Z$ production cross sections    
times leptonic branching ratios in     
$p  p$ collisions at  $14\; \TeV$
using the various MRST parton sets discussed in the text.}
\label{fig:LHCwzdat}
\end{figure}
\newpage

\begin{figure}[H]
\begin{center}     
\epsfig{figure=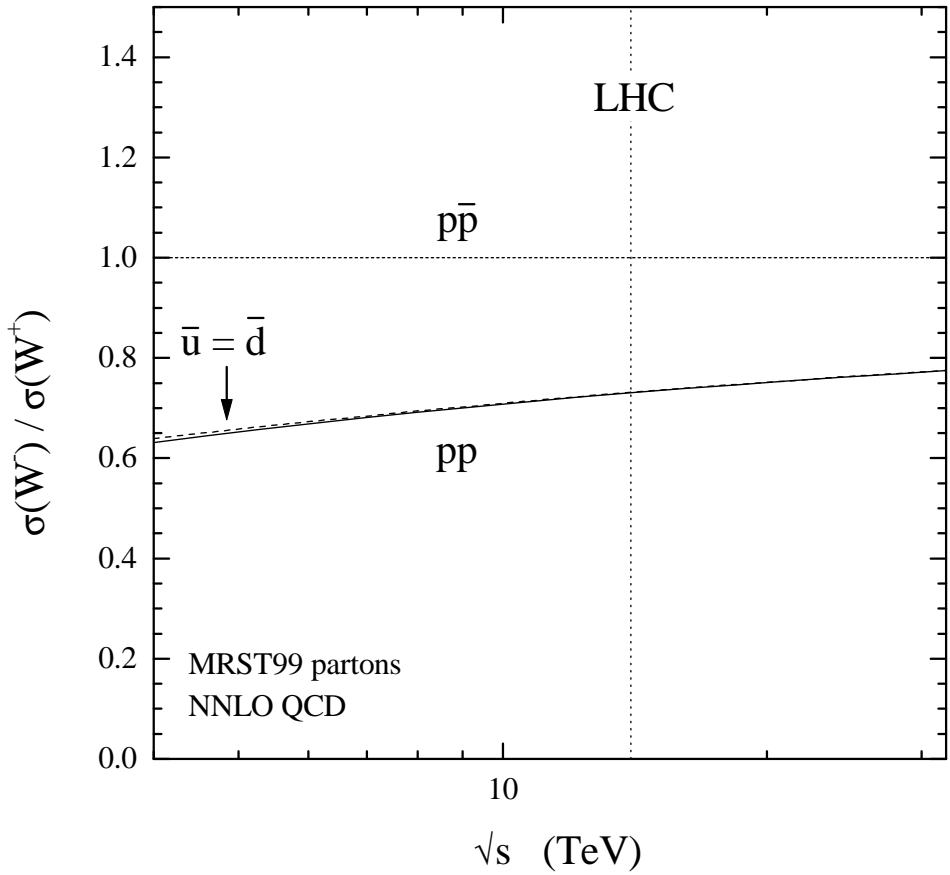,height=20cm}
\end{center}    
\vspace{-1cm}
\caption[]{Prediction for the ratio $R_\mp$ of $W^-$ and $W^+$ total cross
sections in $pp$ collisions,
as a function of the  collider energy $\sqrt s$. The default MRST99
partons are used. For $\pp$ collisions the ratio
is 1. Also shown (dashed line) is the prediction obtained by setting
$\bar u = \bar d$ in the quark sea.}
\label{fig:LHCwrat}
\end{figure}                            
\newpage 
                                                                     
\begin{figure}[H]                            
\begin{center}     
\epsfig{figure=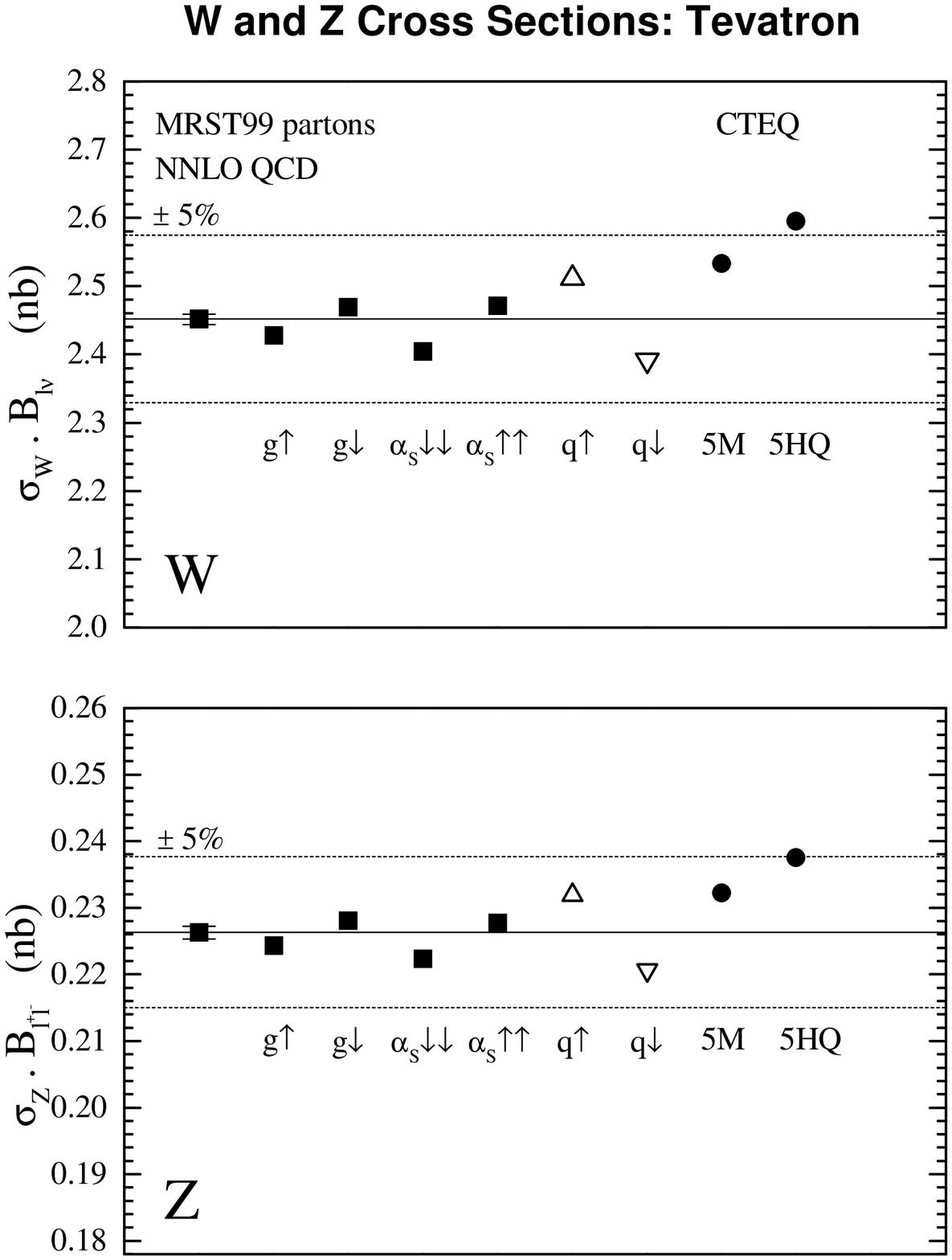,height=20cm}
\end{center}    
\vspace{-2cm}                                                                
\caption{Predictions for the $W$ and $Z$ production cross sections    
times leptonic branching ratios in     
$p \bar p$ collisions at  $1.8\; \TeV$
using the  MRST99, CTEQ5M and CTEQ5HQ \protect\cite{CTEQ5} parton sets.}
\label{fig:TEVwzctq}
\end{figure}                            
\newpage

\begin{figure}[H]
\begin{center}     
\epsfig{figure=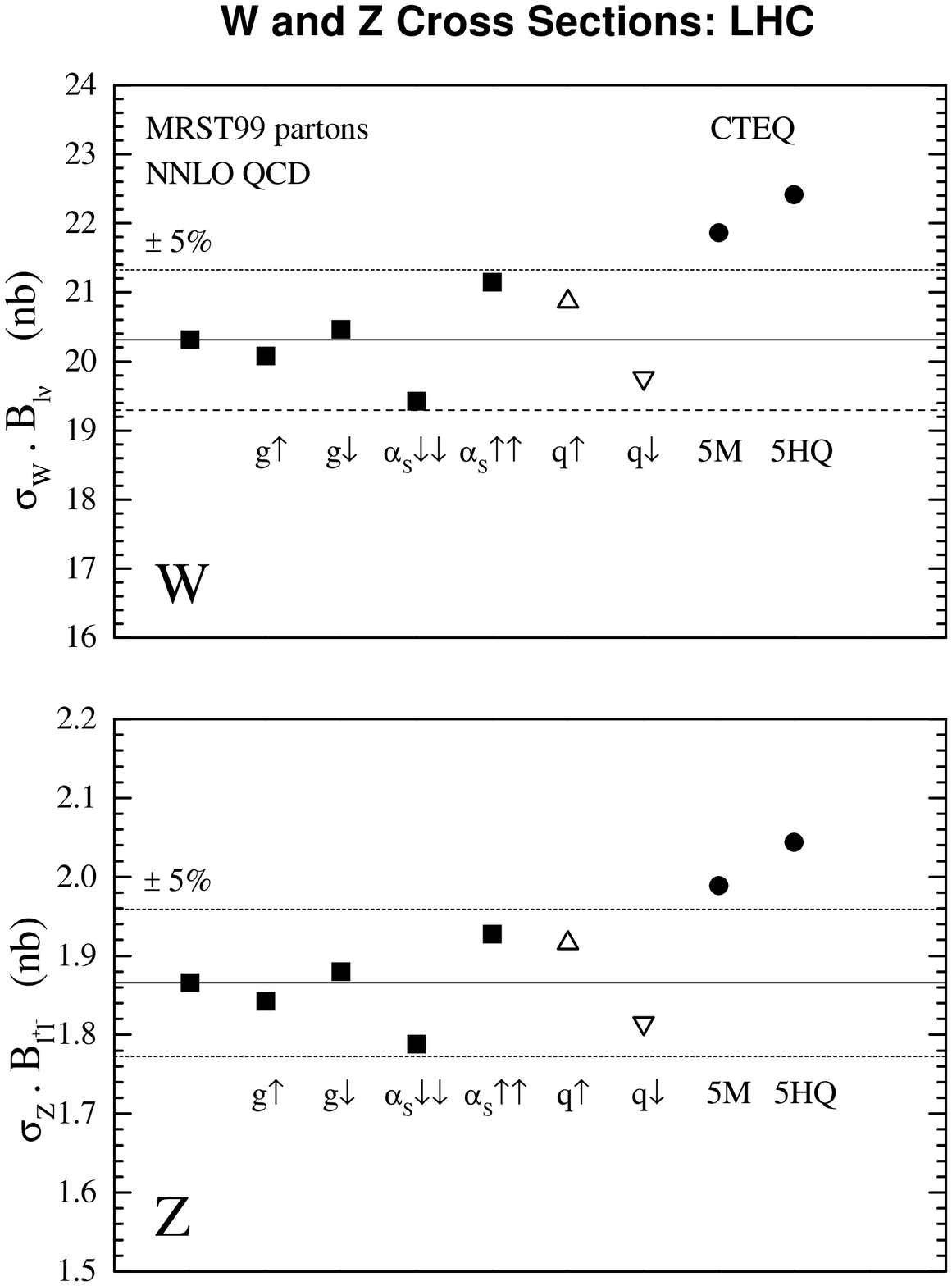,height=20cm}
\end{center}    
\vspace{-2cm}                                                                
\caption{Predictions for the $W$ and $Z$ production cross sections    
times leptonic branching ratios in     
$p  p$ collisions at  $14\; \TeV$
using the  MRST99, CTEQ5M and CTEQ5HQ \protect\cite{CTEQ5} parton sets.}
\label{fig:LHCwzctq}
\end{figure}
\newpage
                                                                     
\begin{figure}[H]
\vspace{-1.5cm}                                                                
\begin{center}     
\epsfig{figure=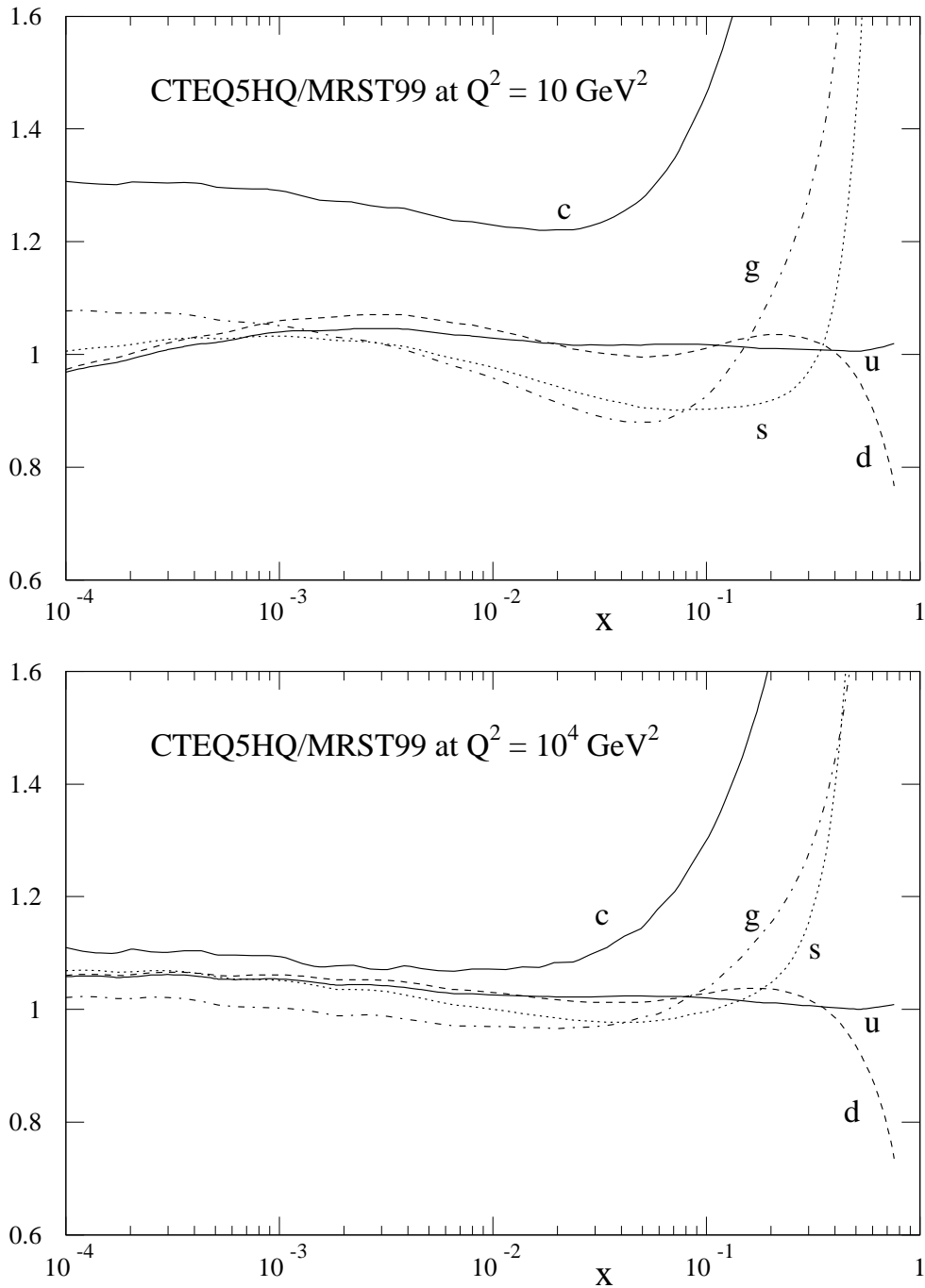,height=20cm}
\end{center}    
\caption{Ratio of the partons of the CTEQ5HQ~\protect\cite{CTEQ5}    
set to those of the default MRST99 set at $Q^2 = 10$ and $10^4$~GeV$^2$.}     
\label{fig:t995hq}
\end{figure}
\newpage

\begin{figure}[H]
\begin{center}     
\epsfig{figure=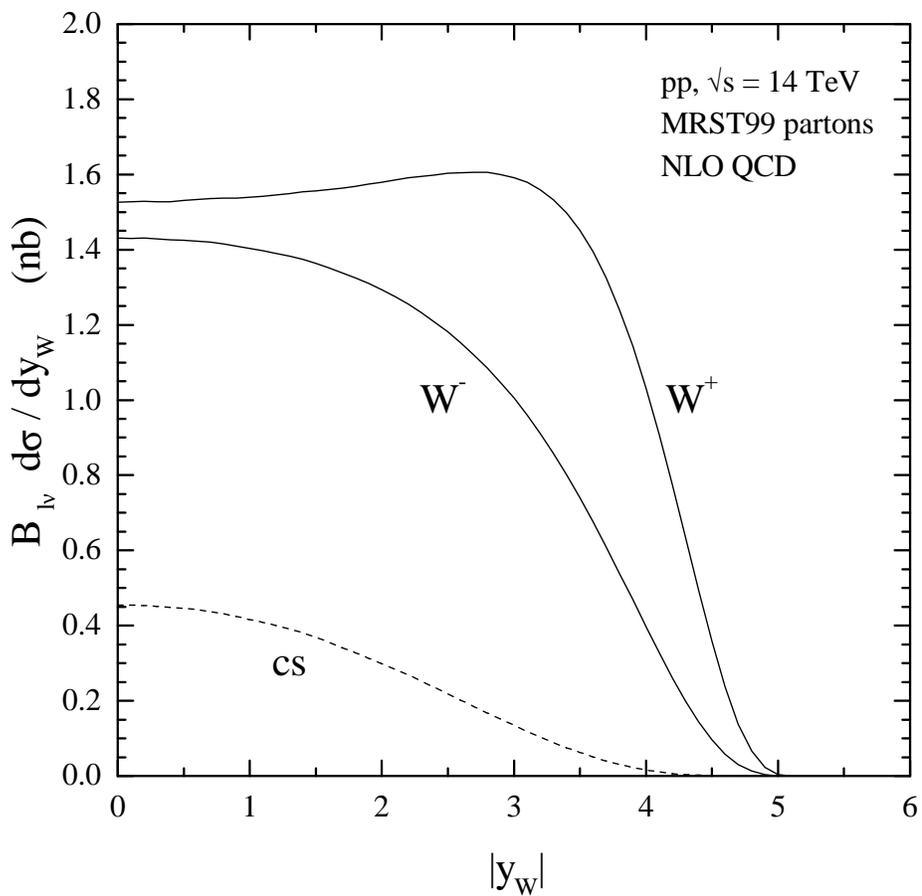,height=20cm}
\end{center}    
\vspace{-2cm}                                                                
\caption{Rapidity distributions of the $W^-$ and $W^+$
bosons at the LHC. Also shown (dashed line) is the (common)
charm--strange scattering contribution.}
\label{fig:LHCwraps}
\end{figure}
\newpage
                                                                     
\begin{figure}[H]
\begin{center}     
\epsfig{figure=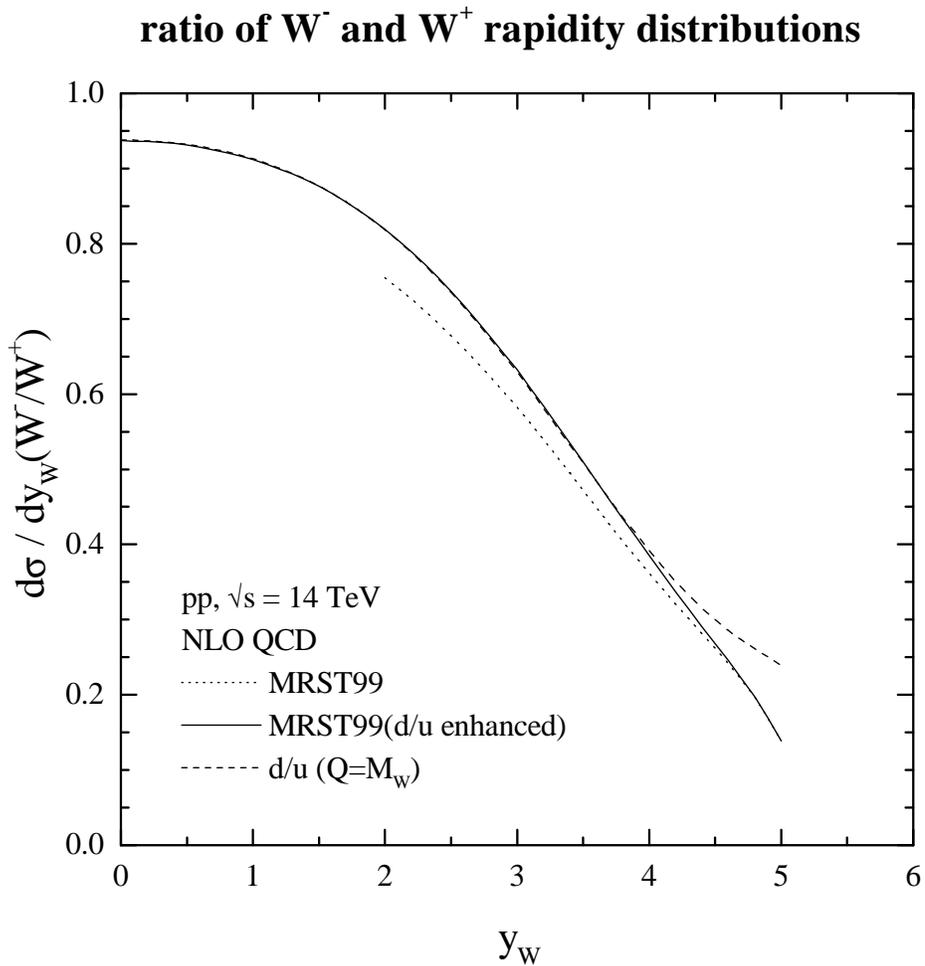,height=20cm}
\end{center}    
\vspace{-2cm}                                                                
\caption{Ratio of the rapidity distributions of the $W^-$ and $W^+$
bosons at the LHC. The solid line corresponds to the default MRST
partons and the dashed line to a modified set (see Table~1) with
an enhanced $d/u$ ratio at large $x$. Also shown (dotted line) 
is the ratio of $d$ to $u$ pdfs evaluated at 
$x=M_W/\sqrt{s} \, \exp(y_W)$
and $Q^2=M_W^2$.}
\label{fig:LHCwraprat}
\end{figure}
\newpage

\begin{figure}[H]
\begin{center}     
\epsfig{figure=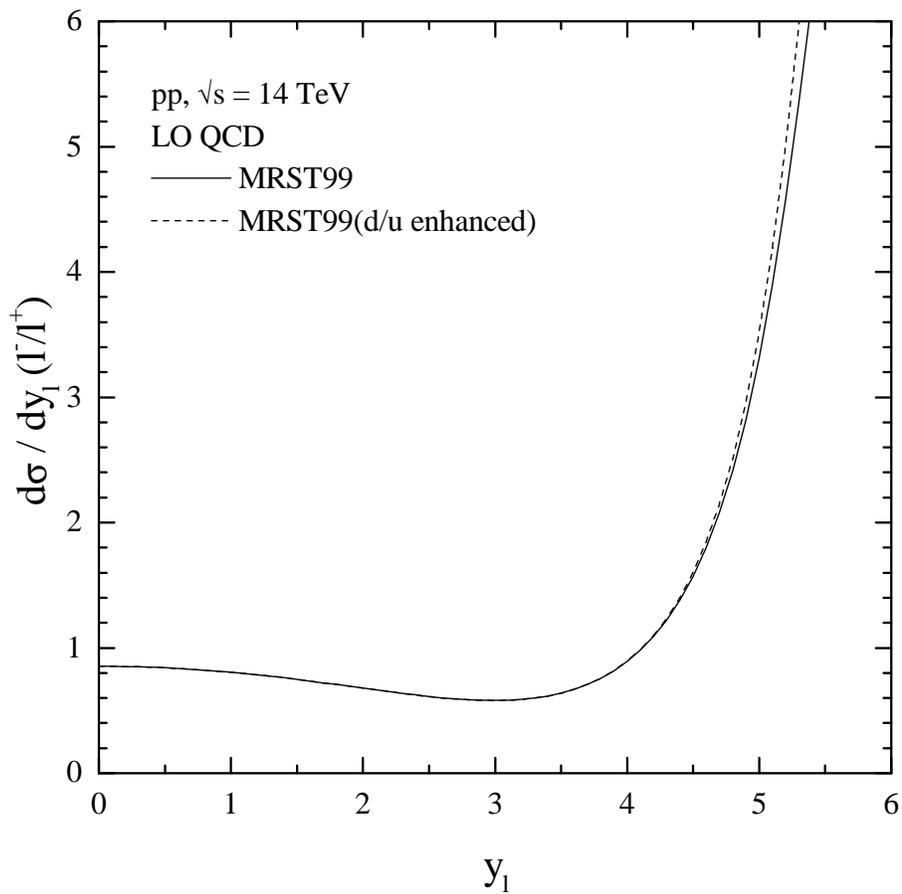,height=20cm}
\end{center}    
\vspace{-2cm}                                                                
\caption{Ratio of the charged lepton 
rapidity distributions from  $W^-$ and $W^+$
production and decay at the LHC. The solid line corresponds to the default MRST
partons and the dashed line to a modified set (see Table~1) with
an enhanced $d/u$ ratio at large $x$.}
\label{fig:LHClraprat}
\end{figure}
\newpage

\begin{figure}[H]
\begin{center}     
\epsfig{figure=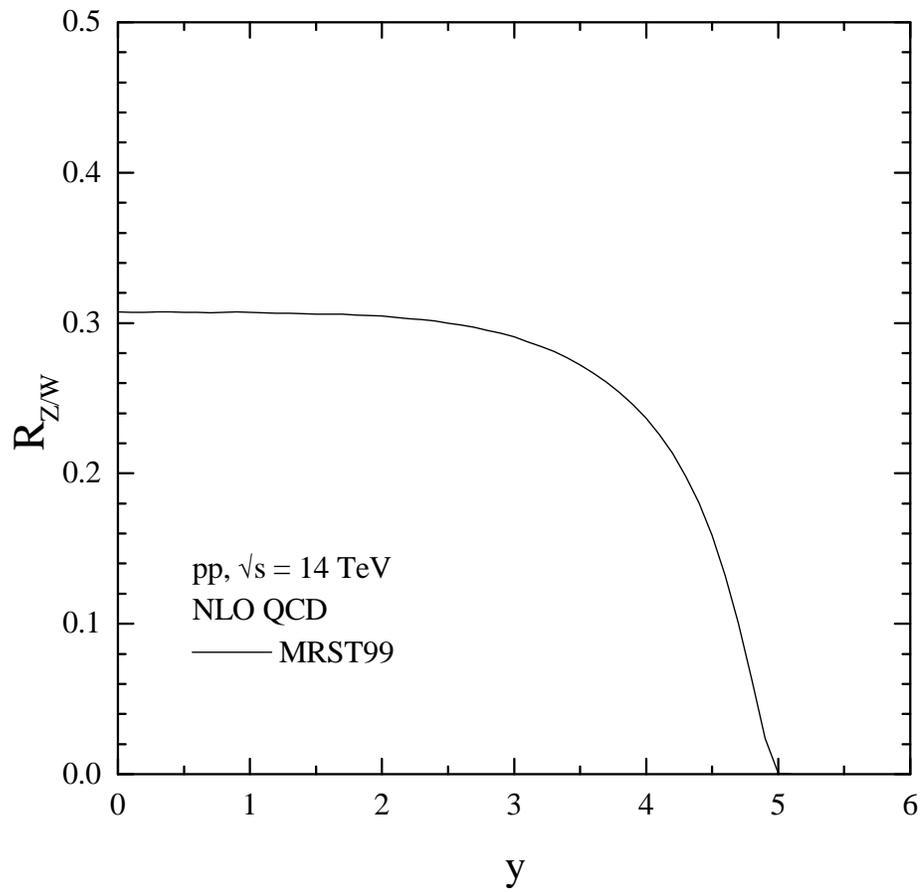,height=20cm}
\end{center}    
\vspace{-2cm}                                                                
\caption{Ratio of the rapidity distributions of the $Z^0$ and $W^\pm$ 
bosons at the LHC, calculated using the  default MRST
partons.}
\label{fig:LHCwzraprat}
\end{figure}
                                                                            
\end{document}